%% file: main.tex
\documentclass[sigconf]{acmart}
\usepackage{tikz}
\usetikzlibrary{positioning, arrows.meta, fit, backgrounds, calc}
\usepackage{listings}
\usepackage{xcolor}
\lstdefinestyle{promptblock}{
  basicstyle=\ttfamily\scriptsize,
  breaklines=true,
  breakatwhitespace=true,
  columns=flexible,
  frame=single,
  rulecolor=\color{gray!40},
  backgroundcolor=\color{gray!5},
  aboveskip=0.5em,
  belowskip=0.5em,
  xleftmargin=1em,
  xrightmargin=1em
}

\AtBeginDocument{%
  }
 
\acmConference[Preprint]{Preprint}{}{}
\renewcommand{\footnotetextcopyrightpermission}[1]{%
  \footnotetext{%
    Preprint.
  }%
}

\begin{document}

\newcommand{\systemname}{Sketch-Based Access Control}
\newcommand{\shortname}{SBAC}
\title{\systemname{}: A Multimodal Interface for Translating User Preferences into Intent-Aligned Policies}


\author{Kyzyl Monteiro}
\email{kyzyl@cmu.edu}
\orcid{0000-0002-2723-9500}
\affiliation{%
  \institution{Carnegie Mellon University}
  \city{Pittsburgh}
  \state{PA}
  \country{USA}
}
\author{Sauvik Das}
\email{sauvik@cmu.edu}
\orcid{0000-0002-9073-8054}
\affiliation{%
  \institution{Carnegie Mellon University}
  \city{Pittsburgh}
  \state{PA}
  \country{USA}
}

\renewcommand{\shortauthors}{Monteiro et al.}

\begin{abstract}
Developing simple and expressive access controls---interfaces to specify policies that define who should have access to resources and under what circumstances---is a longstanding challenge in usable security. 
We present \systemname{} (\shortname{}), a sketch-based, AI-assisted access control authoring system that combines the expressive power of sketching with the interpretive capabilities of multimodal large language models (MLLMs) to support the interpretation and validation of policy specifications as they are iteratively refined. 
Through a formative study with 14 participants, we identified three design requirements and developed a human-AI collaborative workflow composed of three stages---Specify, Analyze, and Test---enabled by the system’s ability to maintain and interpret evolving access control specifications. 
In a user evaluation with 14 participants grounded in their real-world access control scenarios, we found the system and the workflow helped participants progressively refine initially underspecified preferences into more complete and precise policies---surfacing gaps they had not anticipated, resolving ambiguities through dialogue, and validating policy behavior through concrete scenarios.

\end{abstract}



\begin{teaserfigure}
  \includegraphics[width=\textwidth]{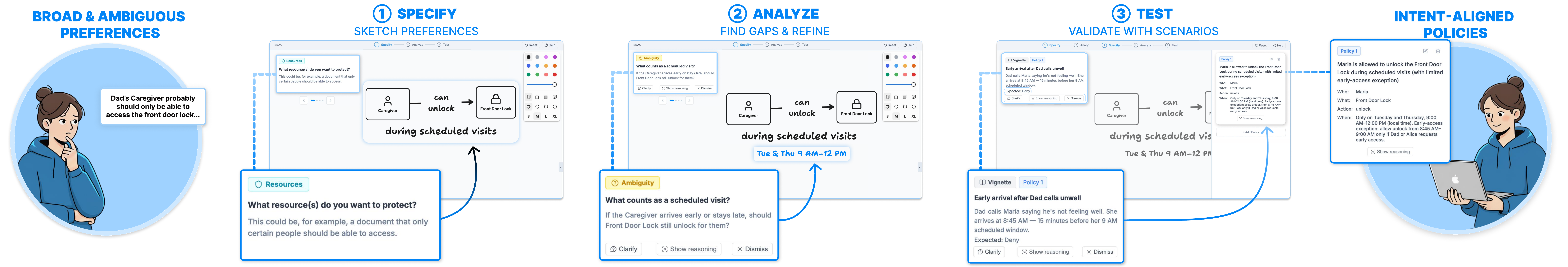}
  \vspace{-20pt}
  \caption{\systemname{} helps users translate broad and ambiguous access control preferences into structured, validated, intent-aligned policies through a three-stage workflow. Users sketch preferences on a freeform canvas where new specifics emerge through the act of drawing (Specify), the system surfaces ambiguities and risks (Analyze), and scenario-based vignettes validate policy behavior at boundary conditions (Test).}
  \Description{A five-part horizontal progression. On the far left, a thought bubble contains scattered words representing a user's loose access control preferences. Three middle panels show the SBAC interface across its three stages: Specify shows a freeform sketch canvas with guidance cards, Analyze shows insight cards flagging an ambiguity with show-reasoning annotations on the sketch, and Test shows a vignette card presenting a boundary-condition scenario. On the far right, a clean structured policy is displayed in labeled fields: Who, What, Action, and When.}
  \label{fig:teaser}
\end{teaserfigure}


\maketitle

\input{sections/1intro}

\input{sections/2threat_model}

\input{sections/3related_work}

\input{sections/4study}

\input{sections/5system_design}

\input{sections/6evaluation}

\input{sections/7discussion}

\input{sections/8conclusion}

\begin{acks}
\end{acks}

\bibliographystyle{ACM-Reference-Format}
\bibliography{references}

\input{sections/appendix}

\input{sections/supmaterial}

\end{document}

%% file: sections/1intro.tex
\section{Introduction}


We must often settle when making access control decisions. If Alice wants her elderly father’s caregiver to unlock his front door only during scheduled visits, a typical smart lock would force her to choose between granting standing access, denying access altogether, or manually providing one-time access for each visit. If Bob wants to share a sensitive document with a colleague only during a meeting, Google Docs offers no obvious mechanism for access to be revoked after one view or one day. The problem is not that such access control policies --- i.e., rules that govern who can access files, devices, and other resources under what circumstances --- \emph{cannot} be supported, but that users lack simple, flexible interfaces for specifying preference-aligned policies. Indeed, creating a simple, flexible access control interface remains one of the longest standing challenges in usable security.


Simple, flexible access control interfaces are challenging to build because users reason about access control in terms of broad, loosely structured \textit{preferences}~\cite{bauer2009real, he2018rethinking} that are often incomplete and ambiguous (should Bob's colleague be able to make a personal copy of the document?). Yet, to become enforceable policies, these preferences must be translated into specific, structured rules.
Therein lies the disconnect: access control interfaces require users to create specific and structured rules, but users do not think about access control in those terms.

Unsurprisingly, prior work has shown that users struggle to articulate their preferences in the language and abstractions required by formal, structured access control interfaces~\cite{reeder2007usability, wumodeling, madejski2012study, mazurek2010access, he2018rethinking}.
Thus, today's interfaces fall at two extremes: simple but coarse controls (e.g., ``allow once'' vs. ``allow always'' permissions) that allow users to make simple decisions that may not be fully aligned with their preferences, or flexible but complex interfaces (e.g., chmod) that require users to invest substantial effort to create detailed but error-prone policies.

What if users could sketch their preferences? Sketching is promising because it helps users externalize loose ideas that are difficult to express through structured input. In usable security and privacy, sketching has surfaced nuanced mental models of access and data flow~\cite{kang2015my, oates2018turtles}; more broadly in HCI, sketching has served as an expressive input mechanism for specifying structured output across domains~\cite{namgyal2023midi, shah2010robust, skubic2007using, chung2022talebrush, montana2021sketching}.
Critically, sketching offers affordances that other input mechanisms alone do not: a sketch canvas is \textit{persistent and spatial}---elements remain visible and revisable across the entire interaction, and users must commit to a structure that makes preferences explicit, which in turn forces them to reason about nuances they might otherwise leave implicit.

However, sketching as an \emph{input} mechanism has long been challenging because freeform sketches are highly abstract and variable~\cite{eitz2012humans, alvarado2007sketchread}.
We hypothesize that advances in Multimodal Large Language Models (MLLMs)~\cite{hurst2024gpt} now make it feasible to compile sketches of access control \textit{preferences} into structured access control \textit{policies}.


In this paper, we introduce \systemname{} (\shortname{}), a sketch-based, AI-assisted access control authoring system.

\shortname{} enables \textit{expressive flexibility}---letting users articulate nuanced access control preferences through sketch and natural language on their own terms---and \textit{simplicity}---automatically translating those expressions into structured policies. It achieves this simplicity and flexibility through a three-phase workflow that allows users to iteratively specify their preferences using natural language and sketches, analyze how those preferences might be converted into structured rules and surface ambiguities and risks, and then test their crafted policies against generated boundary-condition vignettes (see Figure~\ref{fig:teaser}).

We designed \shortname{} using an experience prototyping approach~\cite{buchenau2000experience}. We started by developing an initial technology probe with which we ran a formative study with 14 participants, who helped us elicit key design requirements. To examine how \shortname{} supports policy authoring in practice, we then conducted a summative evaluation with a separate cohort of 14 participants grounded in their own real-world access control scenarios. These participants found the system simple and flexible, crafting final policies that closely reflected their true preferences. Notably, the workflow did not merely capture pre-existing preferences---it helped participants form and refine them through iterative articulation, analysis, and testing. Sketching prompted users to consider new relationships and constraints on their preferences; system-generated feedback made both the user's preferences and the system's interpretation mutually visible, surfacing gaps neither side had identified alone; and, scenario-based testing enabled participants to confirm that their policies aligned with their intent and make informed decisions about what risks they might accept. Participants experienced this process as collaborative co-authorship, producing more complete and precise policies than either user or system would have reached alone. The value of \shortname{} lay in providing a sketch-based input mechanism, and more importantly, in structuring access control authoring as a staged process of expression, interpretation, analysis, and validation.

In short, we contribute \shortname{} --- a simple and flexible multimodal interface that helps users translate ambiguous security and privacy preferences into structured, intent-aligned access control policies.




%% file: sections/2threat_model.tex

%% file: sections/3related_work.tex
\section{Related Work}
\input{figures/positioning-figures}

\subsection{Access Control Specification}
Access control research has long emphasized \textit{models} for representing and enforcing policy, including Role-Based Access Control, Attribute-Based Access Control, and context-aware approaches such as World-Driven Access Control~\cite{sandhu1998role, hu2015attribute, roesner2014world}, but comparatively less work addresses how end users \textit{specify} policies. Prior studies show that users struggle to translate social, situational preferences into rigid policy abstractions, leading to errors, misalignment, and regret~\cite{madejski2012study, mazurek2010access, he2018rethinking, wumodeling}. Visual and template-based systems improve comprehension through matrices, structured visualizations, and policy patterns~\cite{reeder2008expandable, morisset2018building, johnson2010usable}, while natural-language approaches accept informal descriptions and generate structured policy representations, with some additionally checking for structural properties such as conflict-freedom and conformance~\cite{slankas2013access, campagna2018controlling, taninaka2025transparent, cheng2025say}. Separate lines of work support downstream checking through conflict detection, formal verification, and embedded policy analysis~\cite{reeder2009effects, fisler2005verification, hughes2008automated, davy2008policy}. Across these lines of work, existing approaches focus on structural correctness---whether policies are conflict-free, conformant, or faithfully translated---rather than helping users surface what they may have failed to consider. \shortname{} builds on these threads by combining freeform specification, user-facing normative analysis (i.e., surfacing the downstream impact of specification choices by evaluating whether policies are appropriate for the social context), and scenario-based validation in a single end-user authoring workflow.

\subsection{Sketching for Capturing User Intent in Security and Privacy}
Sketching has proven valuable in usable security and privacy as a way to externalize mental models that structured interviews often miss: prior work uses drawing to surface how people conceptualize privacy, data flow, and network architecture~\cite{kang2015my, oates2018turtles}. More broadly, Buxton argues that sketching is generative, helping people form ideas rather than merely record them~\cite{buxton2010sketching}. Recent work has begun moving sketching toward a more active role in privacy design: Wen et al.\ show that heuristic-guided sketching helps novice data scientists produce more complete and readable privacy sketches~\cite{wen2025teaching}. Yet sketching has remained mostly a \textit{research or communication instrument} because systems could not easily interpret ambiguous visual input. Recent multimodal large language models make that interpretation increasingly feasible~\cite{hurst2024gpt}. \shortname{} extends this line of work by turning sketching from an elicitation method into a live interface for authoring access control policies.

\subsection{Sketch-Based and Canvas-Based Human-AI Co-Creation}
Human-AI co-creation research shows that a canvas can function both as a thinking space for human thought and as a shared medium for interaction with AI. Intent Tagging uses canvas-based intent tags to steer generation and help users reflect on what they want~\cite{gmeiner2025intent}; ImaginationVellum shows that spatial arrangement and freeform strokes shape generative output, making the canvas itself part of the prompt~\cite{marquardt2025imaginationvellum}. Beyond creative domains, Code Shaping demonstrates that freeform sketching can serve as an expressive input mechanism for structured output: programmers sketch directly on code to guide AI-driven edits~\cite{yen2025code}, though these systems assume user intent is relatively well formed and focus on interpretation fidelity. SketchDynamics moves closer to our setting: building on the observation that ambiguity can be productive in design~\cite{gaver2003ambiguity}, it shows that intent emerges through interaction rather than existing fully formed at the outset~\cite{li2026sketchdynamics}. Across this body of work, canvas-based co-creation systems help users discover and articulate what they want---but do not evaluate whether the resulting specification satisfies requirements the user may not have considered. In access control, such gaps can have direct security and privacy consequences~\cite{madejski2012study}. \shortname{} extends canvas-based co-creation into this high-stakes setting by pairing sketch-based preference articulation with explicit normative analysis and scenario-based testing, surfacing gaps in contextual appropriateness during specification rather than after deployment.

%% file: figures/positioning-figures.tex
\definecolor{sbaclight}{HTML}{9FD2FF}
\definecolor{sbacdark}{HTML}{0184F6}

\begin{figure}[t]
\centering
\begin{minipage}[t]{0.46\columnwidth}
\centering
\begin{tikzpicture}[
    scale=0.5,
    every node/.style={transform shape},
]
\draw[gray!40, thick] (0,0) rectangle (7,6);
\draw[gray!40] (3.5,0) -- (3.5,6);
\draw[gray!40] (0,3) -- (7,3);

\fill[sbaclight!25] (3.5,3) rectangle (7,6);

\node[font=\small, text=gray!70, anchor=south, rotate=90, align=center] at (-0.2, 1.5) {Pre-formed\\preferences assumed};
\node[font=\small, text=sbacdark, anchor=south, rotate=90, align=center] at (-0.2, 4.5) {Preference\\formation supported};
\node[font=\small\bfseries, text=black!80, anchor=south, rotate=90, align=center] at (-1, 3) {Surfaces unformed preferences};

\node[font=\small, text=gray!70, anchor=north] at (1.75, -0.1) {No gap surfacing};
\node[font=\small, text=sbacdark, anchor=north] at (5.25, -0.1) {System surfaces gaps};
\node[font=\small\bfseries, text=black!80, anchor=north, align=center] at (3.5, -0.5) {System-driven gap surfacing};

\node[font=\small\bfseries, text=black!70, align=center] at (1.75, 5.0) {Canvas-Based\\Co-Creation};
\node[font=\small, text=black!60, align=center] at (1.75, 4.0) {SketchDynamics~\cite{li2026sketchdynamics}\\ImaginationVellum~\cite{marquardt2025imaginationvellum}\\Intent Tagging~\cite{gmeiner2025intent}};

\node[font=\small\bfseries, text=black!70, align=center] at (5.25, 2.3) {Visual AC\\Editors};
\node[font=\scriptsize, text=gray!60, align=center] at (5.25, 1.7) {(structural gaps only)};
\node[font=\small, text=black!60, align=center] at (5.25, 1.0) {Expandable Grids~\cite{reeder2008expandable}\\VisABAC~\cite{morisset2018building}};

\node[font=\small, text=gray!30] at (1.75, 1.5) {---};

\node[font=\normalsize\bfseries, text=sbacdark] at (5.25, 4.8) {Our Focus};
\node[font=\small, text=sbacdark] at (5.25, 4.4) {SBAC};

\end{tikzpicture}
\\[3pt]
{\scriptsize (a) Visual intent specification space}
\end{minipage}
\hfill
\begin{minipage}[t]{0.50\columnwidth}
\centering
\begin{tikzpicture}[
    scale=0.5,
    every node/.style={transform shape},
]
\draw[gray!40, thick] (0,0) rectangle (7,6);
\draw[gray!40] (3.5,0) -- (3.5,6);
\draw[gray!40] (0,2) -- (7,2);
\draw[gray!40] (0,4) -- (7,4);

\fill[sbaclight!25] (3.5,4) rectangle (7,6);

\node[font=\small, text=gray!70, anchor=south, rotate=90] at (-0.2, 1) {None};
\node[font=\small, text=gray!70, anchor=south, rotate=90] at (-0.2, 3) {Structural};
\node[font=\small, text=sbacdark, anchor=south, rotate=90, align=center] at (-0.2, 5) {Structural +\\Contextual};
\node[font=\small\bfseries, text=black!80, anchor=south, rotate=90, align=center] at (-0.8, 3) {Specification-time analysis};

\node[font=\small, text=gray!70, anchor=north] at (1.75, -0.1) {Rigid};
\node[font=\small, text=sbacdark, anchor=north] at (5.25, -0.1) {Freeform};
\node[font=\small\bfseries, text=black!80, anchor=north] at (3.5, -0.5) {Input expressiveness};

\node[font=\small\bfseries, text=black!70, align=center] at (1.75, 1.4) {Basic AC\\Interfaces};
\node[font=\small, text=black!60, align=center] at (1.75, 0.6) {Permission toggles};

\node[font=\small\bfseries, text=black!70, align=center] at (5.25, 1.4) {Guided Dialogue};
\node[font=\small, text=black!60, align=center] at (5.25, 0.7) {Transparent Barriers~\cite{taninaka2025transparent}};

\node[font=\small\bfseries, text=black!70, align=center] at (1.75, 3.5) {AC Editors +\\Checkers};
\node[font=\small, text=black!60, align=center] at (1.75, 2.6) {\texttt{chmod}, XACML\\Expandable Grids~\cite{reeder2008expandable}\\VisABAC~\cite{morisset2018building}};

\node[font=\small\bfseries, text=black!70, align=center] at (5.25, 3.5) {NL +\\Structural Checks};
\node[font=\small, text=black!60, align=center] at (5.25, 2.6) {Almond~\cite{campagna2018controlling}\\LACE~\cite{cheng2025say}};

\node[font=\small, text=gray!30] at (1.75, 5) {---};

\node[font=\normalsize\bfseries, text=sbacdark] at (5.25, 5) {Our Focus};
\node[font=\small, text=sbacdark] at (5.25, 4.6) {SBAC};

\end{tikzpicture}
\\[3pt]
{\scriptsize (b) Access Control (AC) specification space}
\end{minipage}
\vspace{-10pt}

\caption{Positioning \systemname{} relative to (a)~co-creation and verification systems and (b)~access control specification interfaces. In~(a), \systemname{} combines preference formation support with domain gap detection. In~(b), \shortname{} pairs freeform input with specification-time contextual analysis---evaluating policies against the social norms of the user's scenario.}
\Description{Two positioning grids side by side showing how this work relates to prior approaches. Left (2x2): visual AC editors surface structural gaps but assume pre-formed preferences, canvas-based co-creation systems help users form preferences but do not surface gaps, and this work supports both. Right (2x3): existing AC interfaces vary in input expressiveness and specification-time analysis, and this work pairs freeform input with structural and contextual analysis.}
\label{fig:positioning}
\vspace{-10pt}
\end{figure}
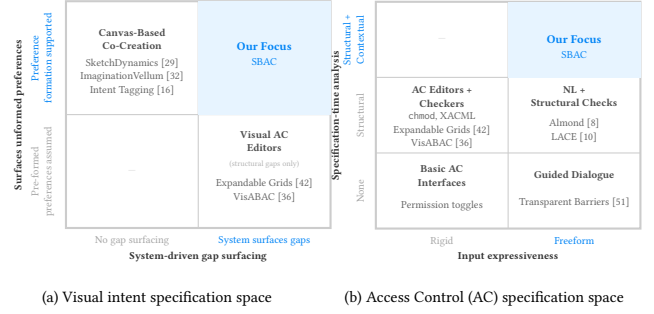

%% file: sections/4study.tex
\begin{figure*}
    \centering
    \includegraphics[width=\linewidth]{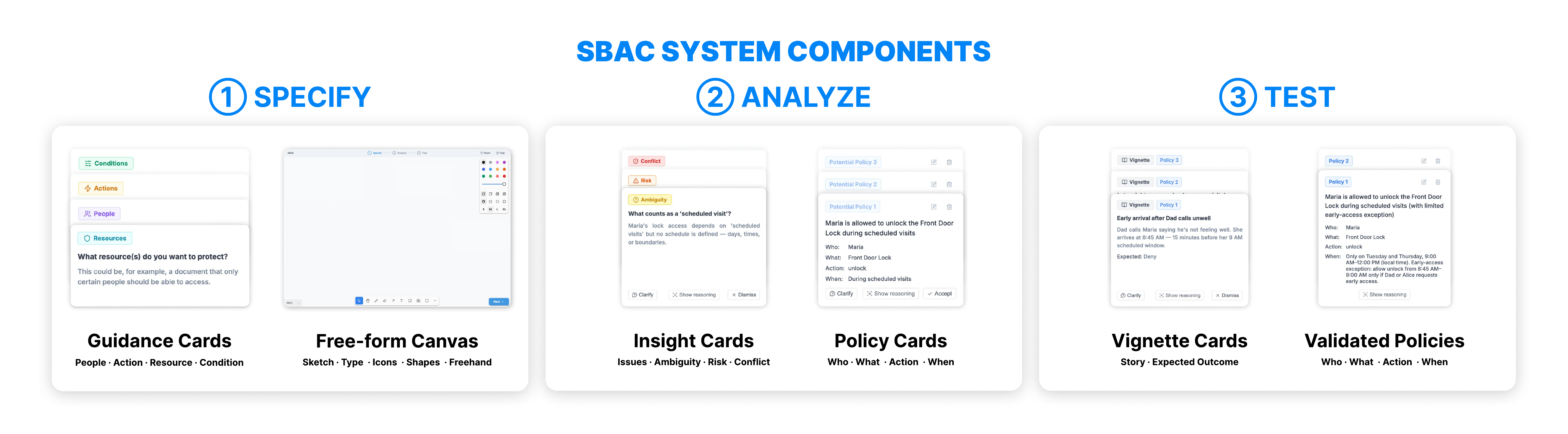}
    \vspace{-25pt}
    \caption{\shortname{} user-facing components: Each workflow stage surfaces card-based artifacts---guidance cards scaffold preference elicitation, insight and policy cards support analysis and refinement, and vignette cards enable scenario-based validation.}
    \Description{Six user-facing components arranged across three workflow stages. Specify shows guidance cards listing People, Action, Resource, and Condition alongside a freeform drawing canvas. Analyze shows an ambiguity insight card and a policy card with Who, What, Action, and When fields. Test shows vignette cards with scenario descriptions and expected outcomes alongside a validated policy card with refined fields.}
    \vspace{-10pt}
    \label{fig:components-flow}
\end{figure*}

\section{Formative Study and Design Requirements}
\label{sec:formative-study}

We used an experience prototyping approach \cite{buchenau2000experience} to elicit concrete design requirements for \systemname{} before building the full system. Specifically, we developed an initial sketch-to-policy technology probe \cite{hutchinson2003technology} and studied how people used it to author access control preferences, where the workflow broke down, and what kinds of support they needed.

\subsection{Method Overview}

We conducted an in-person formative study with 14 participants using a simulated smart office setting populated with devices such as cameras, microphones, badge readers, smart speakers, and environmental sensors. We chose smart environments as our evaluation context because they are where non-expert end users most commonly encounter access control decisions, and where prior work has most extensively documented the gap between users' informal preferences and the structured policies systems require~\cite{he2018rethinking, mazurek2010access, naeini2017privacy}. Participants used the prototype in which they sketched access control preferences on a canvas, invoked an MLLM-based interpretation step, and reviewed the resulting structured policies organized around subject, resource, action, and context~\cite{hu2013guide, nissenbaum2004privacy}. Sessions included two authoring tasks, prompt cards that introduced device capabilities and possible misuse cases, and a post-task interview; we analyzed the resulting sketches and transcripts qualitatively to identify recurring breakdowns and desired forms of support using an inductive thematic analysis \cite{braun2006using}. The study was approved by an IRB, and participants were compensated 15 USD for their time.  Full participant, prototype, procedure, and analysis details are provided in Appendix~\ref{app:formative-study}.

\subsection{Findings and Design Requirements}

Participants found sketching to be expressive and often easier than conventional permission interfaces, but the study also showed that a simple sketch-to-policy pipeline was insufficient. Three recurring findings directly motivated the design requirements that structure \shortname{}.

\paragraph{\textbf{DR1: Scaffold preference formation and articulation.}}
Participants struggled with expressing policies and with deciding what those policies should be. This difficulty stemmed partly from the domain itself---articulating who should access what, when, and under what conditions is cognitively demanding~\cite{reeder2007usability}---and partly from the open-endedness of a blank sketch canvas. As one participant put it, ``It’s kind of hard for me to think of like scenarios when what kind of data should be shared with who\ldots{} that isn’t really a thing I commonly think about’’ (P13). In practice, prompt cards and examples were essential: they helped participants generate more concrete, complete, and nuanced preferences than they would have produced unaided, consistent with prior work showing that guided methods improve specification quality~\cite{karat2006evaluating}. This finding led to \textbf{DR1}: the system should scaffold preference formation and articulation rather than assume users arrive with fully formed specifications.

\paragraph{\textbf{DR2: Surface gaps and divergences in expressed specifications.}}
Participants also needed help recognizing conflicts, risks, and ambiguities in the policies produced from their sketches---both interpretation ambiguity (the system guessed incorrectly and users wanted to understand why) and policy-level concerns (conflicts, overbroad permissions, or privacy risks the user had not considered). One participant explicitly asked for this kind of intervention: ``if the system identifies this language is vague, instead of just guessing it would tell you this language is being vague'' (P1). This finding mirrors prior work suggesting that users miss errors during policy authoring~\cite{madejski2012study}, that rule conflicts can remain hidden~\cite{reeder2009effects}, and that analysis should be embedded into authoring workflows~\cite{davy2008policy}. These findings led to \textbf{DR2}: the system should proactively surface ambiguities, conflicts, and potential risks, surfacing divergences between user intent and system interpretation and gaps in the specification itself.

\paragraph{\textbf{DR3: Validate specifications through concrete scenarios.}}
Participants’ access control preferences were not static: they evolved as participants encountered specific situations, and initial policies consistently revealed gaps when confronted with concrete scenarios. One participant noted that policies ``would change from time to time’’ based on ``how the systems are being used’’ (P8). These gaps surfaced primarily through scenario exposure: prompt cards led participants to discover overlooked concerns---such as privacy implications of thermostats (P7), microphones (P14), and recording devices (P6), among others---and even seeing the system’s interpretation triggered new preferences (P10). This pattern is consistent with prior work showing that access control preferences are context-dependent and difficult to anticipate in the abstract~\cite{mazurek2010access, he2018rethinking, wumodeling}. This finding led to \textbf{DR3}: the system should support scenario-based validation so users can inspect how specifications behave against concrete situations rather than abstract rules alone.

The study also surfaced two cross-cutting concerns that informed the system architecture: coherence across input modes and interpretability of system reasoning. We treat these as design principles that shape the implementation of \shortname{} rather than as standalone requirements.
Appendix~\ref{app:formative-study} provides additional detail and representative participant feedback.

%% file: sections/5system_design.tex
\section{\systemname{}}

To address the design requirements we identified in the formative study, we present \systemname{} (\shortname{}): a sketch-based, AI-assisted access control authoring system. \shortname{} implements a human-AI teaming approach to policy specification. With \shortname{}, users express access control preferences through freeform sketching on a canvas, while the system ``compiles'' those sketches into structured policies, evaluates whether those policies are appropriate for the user's social context using contextual integrity~\cite{nissenbaum2004privacy}, checks for structural issues such as conflicts, and generates concrete scenarios to stress-test policies around boundary conditions. The system walks users through a three-phase workflow---Specify, Analyze, and Test---that progressively moves from preference elicitation to policy interpretation to behavioral validation.

Two cross-cutting concerns from the formative study informed the architecture throughout. First, \textit{coherence}: changes made in one modality---sketch, conversation text, or policy text---must be reflected in the others. Second, \textit{interpretability}: users must be able to trace how the system translated their input into policies and why it flagged potential issues. Both of these concerns are addressed by a shared visual reference system, described in Section~\ref{sec:sketch-interpretation}, that connects sketch elements, interpreted policies, and surfaced insights through a common set of labeled entities.

\begin{figure*}
    \centering
    \includegraphics[width=\linewidth]{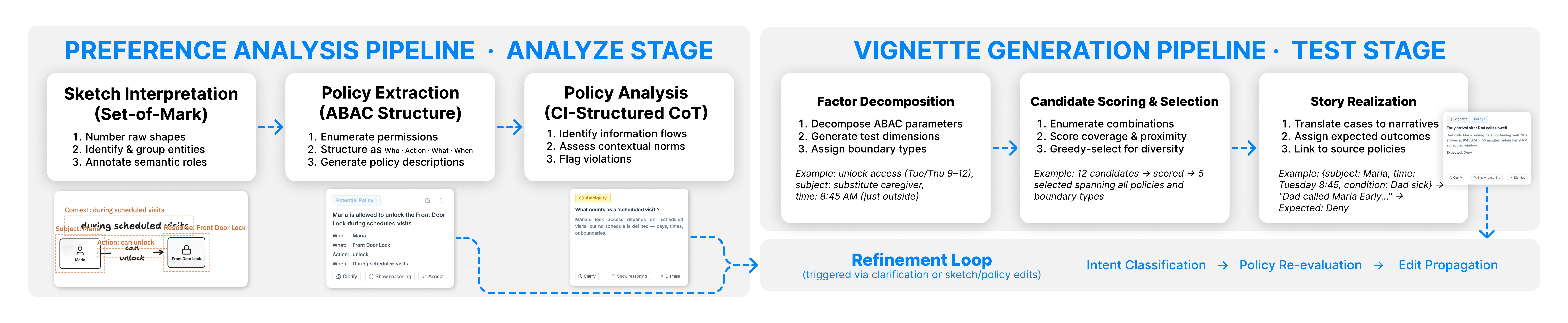}
    \vspace{-25pt}
    \caption{\shortname{} processing pipeline. Preference Analysis interprets a sketch into structured policies and surfaces issues via CI-structured reasoning. A refinement loop handles user clarifications and propagates edits across modalities. Vignette Generation decomposes policies into testable dimensions, selects boundary-probing candidates algorithmically, and realizes them as narrative scenarios.}
    \Description{A horizontal pipeline diagram with two main groups. The Preference Analysis pipeline (Analyze stage) contains Sketch Interpretation, Policy Extraction, and Policy Analysis boxes with output artifacts (SoM-annotated sketch, policy card, insight card). A Refinement Loop below handles user clarifications via intent classification, policy re-evaluation, and edit propagation. The Vignette Generation pipeline (Test stage) contains Factor Decomposition, Candidate Scoring and Selection, and Story Realization boxes with a vignette card output.}
    \label{fig:pipeline}
    \vspace{-15pt}
\end{figure*}

\subsection{User Workflow}

\subsubsection{Phase 1: Specify}
The formative study showed that users need scaffolding to form and articulate access control preferences, given that users do not arrive with fully formed specifications~(DR1). In the Specify phase, a card carousel floating over the canvas displays \textit{guidance cards} that scaffold this elicitation process by introducing the four parameters of an access control policy following the attribute-based access control (ABAC) model (see Section~\ref{sec:policy-structure}): \textit{What} resource(s) do you want to protect? \textit{Who} needs to interact with your resources? What \textit{actions} should each person be able to perform? \textit{When} or under what conditions should these actions be allowed? This sequence grounds preference elicitation in the same structure that will later represent policies. Users express their preferences on a freeform canvas using any combination of the available sketching tools: drawing, handwriting, typed text, icons, or spatial arrangement.

The guidance cards served to frame those preferences in terms of the ABAC parameters rather than to elicit preferences from scratch.
No AI analysis occurs during this phase; the canvas remains a space for freeform expression.

\subsubsection{Phase 2: Analyze}
Users need help recognizing gaps and divergences in their specifications: ambiguities, conflicts, and risks they did not initially consider~(DR2). The Analyze phase addresses this need by first ``compiling'' the sketch into structured access control policies and then evaluating them for potential issues. Each policy is represented in ABAC format using user-facing labels---\textit{Who} (subject), \textit{What} (resource), \textit{Action}, and \textit{When} (context)---and displayed in a policy panel on the right side of the interface. The card carousel, which displayed guidance cards in the Specify phase, now displays \textit{insight cards}, each surfacing a specific risk, ambiguity, or conflict identified in the policy set. Each policy and flagged insight is visually linked to specific sketch elements: toggling the \textit{Show reasoning} button on a policy or issue highlights the corresponding regions on the canvas, showing users \textit{how} the system arrived at each interpretation.

Users can respond to insight cards in three ways: conversationally, directly editing policy fields in the policy panel, or modifying the sketch. Changes in one modality propagate to the others.

\subsubsection{Phase 3: Test}
Specifications evolve as users encounter concrete situations, and initial policies consistently reveal gaps under scenario exposure~(DR3). In the Test phase, the card carousel displays \textit{vignette cards}---short, concrete scenarios that probe the boundaries of the user's policies. Each vignette describes a situation (e.g., ``A guest tries to access the front camera at 11 PM'') and presents an expected outcome (Allow, Deny, or Ambiguous) based on the current policy set. As with insight cards, each vignette is visually linked to the policies and sketch elements it tests, and users can toggle \textit{Show reasoning} to see the corresponding canvas regions. Users review each vignette and can clarify, refine policies, or acknowledge risks (the same interaction patterns available in the Analyze phase). 

The Test phase performs \textit{contextual} validation: it draws on the social roles, resource usage patterns, and situational details of the user's own scenario to generate scenarios that are meaningful precisely because they are grounded in the user's context.

\subsection{Sketch Interpretation via Set-of-Mark Prompting}
\label{sec:sketch-interpretation}

All three phases depend on a shared, multimodal representation that connects sketch elements to compiled policies and flagged issues. A central challenge is that raw canvas primitives do not correspond to meaningful access control entities. A single conceptual entity---such as a person named ``Alice'' or a resource like a camera---can be represented using multiple visual elements combined---for example, an icon, a text label, and a bounding rectangle that groups them together. Alternatively, the entity might be represented more simply with just a label. In contrast, a single shape (such as an arrow) is often used to encode a relationship between two entities. Applying a multimodal LLM directly to the raw canvas would require the model to simultaneously segment the sketch into meaningful regions, classify their roles, and reason about policies, conflating visual parsing with semantic analysis.

We address this challenge through an adaptation of Set-of-Mark (SoM) prompting~\cite{yang2023set} for sketch canvases. In the original SoM technique, visual marks are placed on semantically meaningful image regions (identified by segmentation models) so that an LLM can reference specific regions by their marks. A sketch canvas has no segmentation model---only raw shape data from the drawing library. Our adaptation replaces segmentation with an LLM-driven identification step that consolidates raw shapes into meaningful entities before SoM annotation is applied.

\subsubsection{Mark Identification and Grouping}
When the user enters the Analyze or Test phase, the system overlays numbered marks on every shape and sends paired images to the LLM: the raw sketch to preserve visual fidelity and the numbered version for referencing. The LLM identifies semantically meaningful entities by classifying shapes into ABAC roles, grouping multi-shape composites (e.g., a person icon + ``Alice'' label $\rightarrow$ one entity), and detecting relationships; a typical sketch with 12 raw shapes is consolidated into 3--5 meaningful entities (see Appendix for procedural details).

\subsubsection{Semantic Annotation as Shared Visual Reference}
The identified entities are annotated with role-prefixed labels (``Subject: Alice'', ``Resource: Front Camera'') that serve a dual role: they structure the LLM's input for downstream reasoning, and they provide users with a traceable visual reference through the \textit{Show reasoning} toggle. Because both the system's analysis and the user's verification operate over the same labeled annotations, they function as a shared visual vocabulary. The same entity identifiers flow through analysis, clarification, edit propagation, and vignette generation, enabling the system to maintain a consistent representation across all three stages. Re-identification runs on each phase entry so that the Test phase reasons over entities that reflect any policy or sketch changes made during Analyze.

\subsection{Policy Analysis and Refinement}

The Analyze phase operationalizes DR2 (surfacing gaps and divergences in expressed specifications) through a structured reasoning pipeline and interactive refinement mechanism.

\subsubsection{Policy Structure and Reasoning Framework}
\label{sec:policy-structure}
\shortname{} represents policies using four simplified ABAC parameters---Subject, Action, Resource, Context---presented through an intuitive authoring vocabulary: \textit{Who}, \textit{Action}, \textit{What}, \textit{When}. ABAC was chosen over alternatives (RBAC, DAC, MAC) because its attribute-based parameters accommodate the ad hoc social roles, situational conditions, and diverse resource types that users bring from personal scenarios~\cite{hu2013guide}, and map naturally to how non-experts describe access preferences in plain language~\cite{johnson2010usable}. For the scope of this work, we adopt this template structure as our policy representation, following prior work showing that policies conforming to such templates are implementable and understandable by end users~\cite{johnson2010usable}; the LLM is prompted to extract policies that conform to this format.

While ABAC grounds the policy \textit{structure}, Contextual Integrity (CI)~\cite{nissenbaum2004privacy} provides the evaluative lens the system uses to \textit{analyze} those policies. Each access control policy implies information flows---a camera access policy, for instance, implies a flow of video data from a space to the accessing subject. CI asks whether such flows are appropriate given the social context: the roles involved, the norms governing that setting, and the conditions under which the flow occurs. This enables the system to evaluate not just what the policies \textit{permit}, but whether those permissions are \textit{appropriate}---surfacing risks, ambiguities, and conflicts by assessing policies against the contextual norms of the user's scenario.

\subsubsection{CI-Structured Analysis}
When the Analyze phase begins, the MLLM responsible for analysis receives the SoM-annotated image and structured mark context, and follows a CI-based chain-of-thought reasoning process: (1)~enumerate all permissions as Subject $\rightarrow$ Action $\rightarrow$ Resource [+ Context]; (2)~identify implied information flows; (3)~assess each flow against contextual norms; and (4)~flag violations traceable to specific policies. We adopt this structured prompting approach because Lan et al.~\cite{lan2025contextual} showed that directing LLMs to reason explicitly through contextual parameters before responding substantially reduces inappropriate information disclosure across model families, even without additional training. Because contextual norms are inherently subjective~\cite{nissenbaum2004privacy}, the system surfaces potential concerns for human review rather than claiming correctness.

Each flagged issue is classified as a \textbf{Risk} (permissions violating contextual norms), \textbf{Ambiguity} (underspecified elements), or \textbf{Conflict} (contradictory decisions). These categories are derived from access control verification properties~\cite{hu2017verification}, applied at specification time before any enforcement engine exists. This constitutes structured heuristic assessment informed by NIST and OWASP taxonomies, not formal verification. Each issue carries a structured rationale explaining the information flow, the expected norm, and the gap (see Appendix~\ref{sec:issue-categories-grounding}).

\begin{figure}
    \centering
    \includegraphics[width=\linewidth]{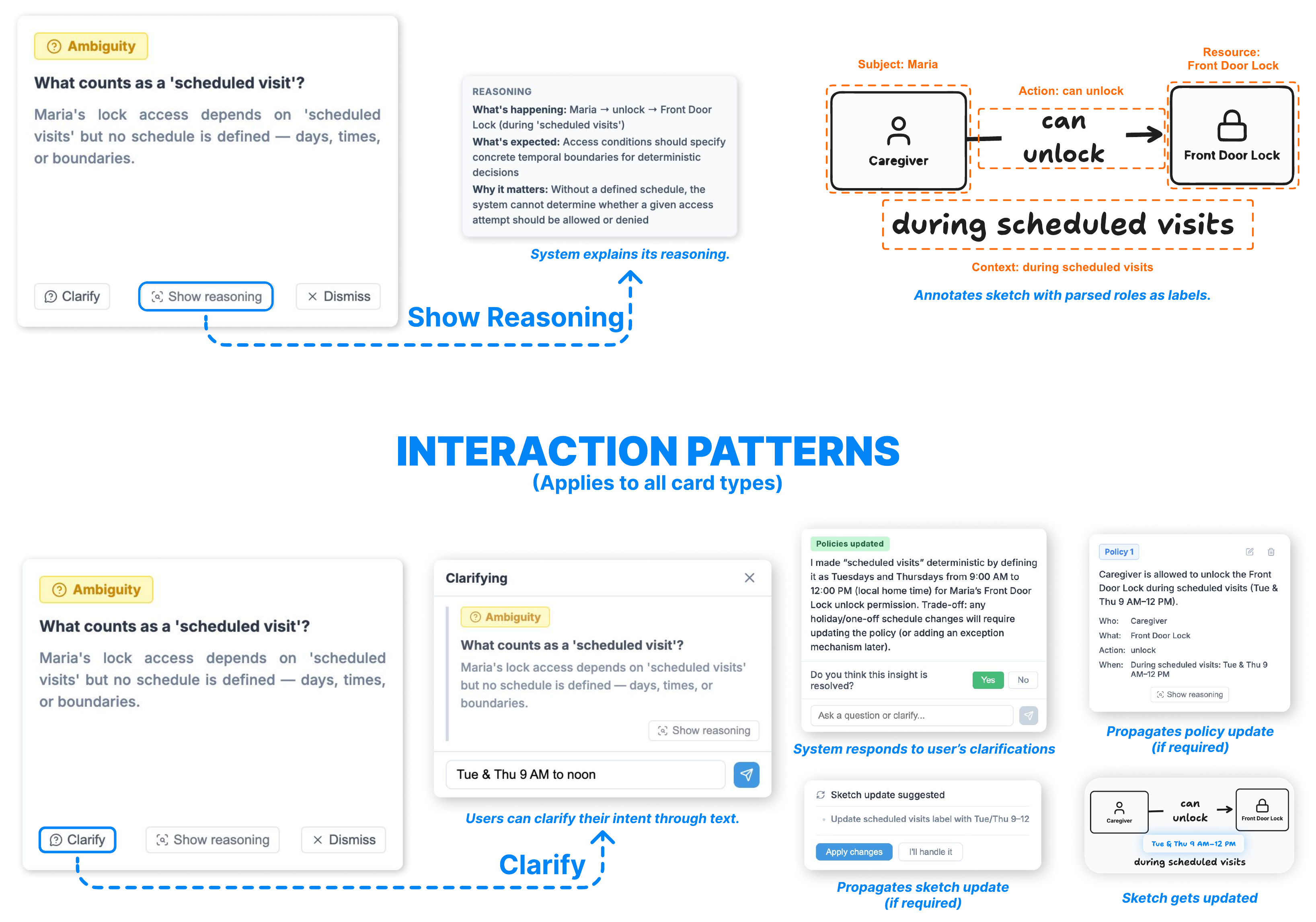}
    \caption{Interaction patterns shared across all card types. \textit{Show Reasoning} traces a system output to the sketch elements it was derived from, displaying the CI-structured rationale and SoM role annotations. \textit{Clarify} lets users respond through conversation; the system classifies intent, updates policies, and propagates changes to the sketch.}
    \Description{Two interaction patterns. Top: Show Reasoning links an ambiguity insight card to a three-part CI rationale (what is happening, what is expected, why it matters) and to the sketch with SoM role annotations (Subject: Maria, Action: can unlock, Resource: Front Door Lock, Context: during scheduled visits). Bottom: Clarify shows a user typing a response in a conversation dialog, the system responding with a policy update and trade-off explanation, the updated policy card, a suggested sketch update with apply or decline options, and the resulting updated sketch.}
    \label{fig:flow}
\end{figure}

\subsubsection{Interactive Refinement and Coherence}
Users can refine their policies conversationally, by editing policies directly, or modifying the sketch. Each action triggers cross-modal propagation to maintain coherence: conversational clarifications are routed through an intent classifier (understand, correct, fix, or explore) that determines whether policy updates are needed, and when a clarification implies the sketch no longer reflects the user's intent, the system can update the sketch itself to maintain alignment; direct policy edits propagate across all referencing policies and insights (see Appendix~\ref{sec:policy-ripple}); and sketch edits trigger incremental re-analysis that incorporates changes alongside the existing clarification history. Propagation is hybrid: simple renames and description edits use deterministic or fast-path updates, while semantically meaningful fixes trigger re-analysis with the full model.

\subsection{Structured Vignette Generation}
Policies that pass structural analysis can still fail in practice: a rule that looks correct in the abstract may produce unintended outcomes when applied to specific people, times, or situations the user did not anticipate. The Test phase addresses this by generating concrete scenario-based vignettes that probe how policies behave at their boundaries. To generate these vignettes, we employ the factorial vignette methodology~\cite{martin2016measuring, naeini2017privacy}, where systematically varying contextual factors across short scenarios isolates which dimensions shape privacy judgments. Adapting this research instrument, \shortname{} generates vignettes from the user's policies by systematically varying policy dimensions to surface gaps and boundary conditions before deployment. 

To identify \emph{where} policy boundaries lie and \emph{which} boundaries to test, and to ensure systematic broad coverage, the vignette generation pipeline has three stages. The pipeline interleaves LLM reasoning with deterministic algorithmic steps: \textit{(1) Factor Decomposition}, the LLM decomposes each policy's ABAC parameters into testable dimensions with boundary types (just inside, just outside, clearly outside, ambiguous); \textit{(2) Candidate Scoring and Selection}, a deterministic algorithm enumerates and scores candidate test cases; and \textit{(3) Story Realization}, the LLM translates selected cases into natural-language vignettes with CI-format rationale---a strictly translational step that does not invent scenarios or alter expected outcomes.

The enumeration generates baselines, single-factor variations, and two-factor combinations from the decomposed dimensions; candidates are scored on boundary proximity, conflict potential, coverage diversity, and novelty, then selected via a greedy diversity-aware algorithm that penalizes redundancy after each pick to avoid trivial or repeated scenarios. Each vignette card is linked to the policies and sketch elements it tests through the shared mark identifiers from sketch interpretation (see Appendix~\ref{sec:scoring-dimensions} for the full pipeline specification and algorithmic details).

\subsection{Implementation}
\shortname{} is implemented as a React web application using TLDraw~\cite{tldraw} for the sketch canvas, with a serverless backend that maintains persistent session state across the three workflow phases. All LLM-powered components use OpenAI models: reasoning-intensive calls (identification, analysis, deep clarification, factor decomposition, story realization) use a frontier model (GPT-5.2), while latency-sensitive calls (intent classification, edit propagation) use a smaller model (GPT-4.1-mini). The system's architecture and prompts are designed to be model-agnostic, though we validated the workflow only with the model family used in this study; we have not tested the quality of outputs with other model families. Prompts were iteratively refined through pilot testing with representative scenarios before the formal study; final prompt specifications are provided in the supplementary material. 

%% file: sections/6evaluation.tex
\section{Evaluation}

We conducted a lab-based user study with 14 participants to answer three research questions about \shortname{}: (RQ1) To what extent is \shortname{} simple and flexible enough for end users to author intent-aligned access control policies? (RQ2) How well does the system address the design requirements identified in the formative study? (RQ3) To what extent does using SBAC affect participants' specified access control preferences? To ground the evaluation in lived experience, we based each participant's study session on a scenario, from their lives, that they already knew well. In the findings below, we first examine whether participants found the workflow simple, flexible, and aligned with their intent (RQ1), then how the workflow supported them across the Specify, Analyze, and Test stages (RQ2), and finally how using SBAC reshaped the access control preferences they articulated (RQ3).

\subsection{Participants}

We recruited 14 participants (8 women, 6 men) aged 18 to 53 through our institution's participant pool and convenience sampling. Most (11/14) reported basic familiarity with permissioning systems and sharing settings; one reported no familiarity (P10), and two had studied or built access control systems (P5, P14). In terms of hands-on experience, participants ranged from everyday permissions such as Google Docs sharing (P1, P4, P7, P8, P12), to advanced settings involving roles or conditions (P2, P6, P10, P11), managing access on behalf of others (P3, P9, P13), and designing or building systems (P5, P14). Scenarios spanned homes, offices, labs, and institutional buildings, involving devices and people participants personally knew (see Appendix~\ref{app:participant-scenarios}).

\subsection{Procedure}

Each study session consisted of four phases and lasted approximately 60 minutes. In the \textit{pre-task phase}, participants described a smart-device scenario and wrote their initial access control preferences in free text.

In the \textit{main task phase}, after a system walkthrough, participants used \shortname{} to work through the Specify, Analyze, and Test stages with their own scenario. In the \textit{post-task phase}, participants re-stated preferences from memory (without access to system output), rated each final policy for alignment with their intent (5-point Likert), and completed the SUS, NASA-TLX, and a custom questionnaire. In the \textit{interview phase}, we conducted a semi-structured interview covering how participants formed preferences, interpreted system outputs, and reasoned about surfaced issues.

\subsection{Data Collection and Analysis}

We administered the SUS, NASA-TLX, a custom 15-item questionnaire aligned with the design requirements, and per-policy intent-alignment ratings (i.e., each participant's perceived alignment between their authored policies and their intent), which together addressed RQ1--RQ3---with RQ1 primarily captured through these instruments, as they directly measure perceived simplicity and flexibility. For RQ2, we additionally recorded interaction logs and policy revisions across all three stages, constructing structured session logs from which we coded 10 refinement patterns (Table~\ref{tab:refinement-types}). For RQ3, we compared pre-task and post-task free-text preferences to check whether refinements persisted in participants' unaided articulations. We thematically analyzed interview transcripts~\cite{braun2019reflecting} across all three RQs, organizing themes around workflow stages and evaluation constructs.

\subsection{Findings}

Overall, participants found \shortname{} simple and flexible enough to author policies that closely matched their intent, the staged workflow addressed the design requirements identified in the formative study, and the process of using SBAC helped participants refine their initially ambiguous access control preferences. The findings reported below---both quantitative and qualitative---draw on our survey instruments (SUS, NASA-TLX, and a custom 15-item questionnaire; 5-point Likert scale; see Appendix~\ref{app:eval-instruments} for item-level breakdowns) and thematic analysis of interviews. Supplementary findings are in Appendix~\ref{app:eval-supplementary}.

\subsubsection{\textbf{RQ1: Participants found \shortname{} to be simple, flexible, and intent-aligned}}

\paragraph{\textbf{\shortname{} helped participants author policies that matched their true preferences.}} For RQ1, the quantitative and qualitative evidence converged. Participants reported strong alignment between their final policies and their intent (Alignment: M=4.71/5; Confidence: M=4.57/5), and gave high policy-level alignment ratings (M=4.85/5), with 10 of 14 participants rating all of their policies at 5/5 (strongly agree). These results suggest that participants saw the resulting policies as faithful to what they actually wanted, rather than settling for acceptable approximations.

\paragraph{\textbf{Participants experienced the workflow as both simple and flexible.}} Participants rated the workflow as simple (M=4.29/5) and flexible (M=4.50/5), and they described it as both usable and manageable rather than burdensome (SUS: M=70.7, SD=15.2; NASA-TLX: M=20.8, SD=11.0). Our interviews reinforced that interpretation. Participants valued having a structured workflow for a task they otherwise found difficult to begin: ``The fact that you have a structured workflow for even thinking about defining these policies is really nice [...] I wouldn't know where to start'' (P3). They also valued the workflow's expressive breadth: ``I like that you could use natural language... I just colored arrows, and it seemed to understand what I meant. But you could do something else entirely'' (P14). RQ2 unpacks which parts of the workflow produced those impressions.

\subsubsection{\textbf{RQ2: \shortname{} addressed the workflow support needs identified in the formative study}}

For RQ2, we organize the findings around the three design requirements identified in the formative study.

\vspace{2mm}
\noindent\textbf{DR1: Preference formation and articulation.}
\paragraph{\textbf{Sketching helped participants form and elaborate their preferences.}} Participants reported that \shortname{} helped them form their access control preferences (M=4.68/5). Of the 14 participants, 11 sketched access control preferences that went beyond what they had written in the \emph{pre-task phase}: they added subjects, actions, conditions, or resources not present in their initial descriptions. The remaining three initially mapped their pre-task preferences exactly but refined them during the Analyze and Test stages. As P6 explained, ``the process of translating my understanding of the policy into a graph, that part can already help me reflect on the relationship between the different parties in the policies.'' Rather than merely transcribing preexisting preferences, the Specify stage helped participants form and elaborate them.

\begin{figure*}[t]
\centering
\includegraphics[width=\textwidth]{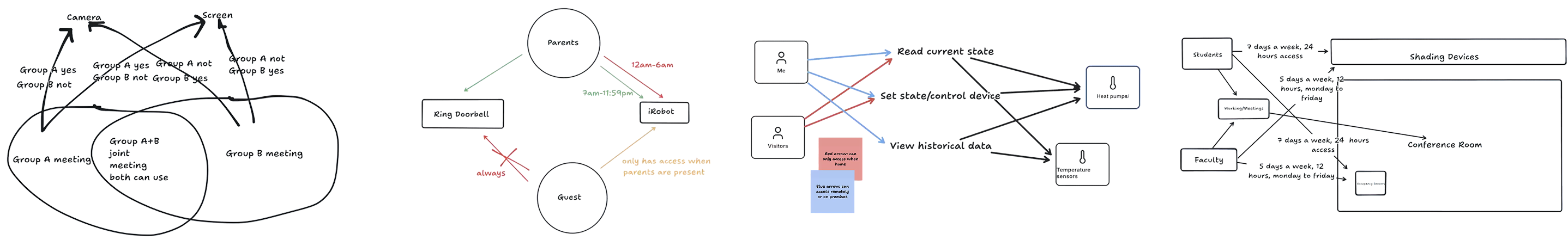}
\vspace{-2pt}
\begin{minipage}[t]{0.25\textwidth}\centering\small(a) Overlapping group permissions\\\scriptsize[Venn diagram]\end{minipage}%
\begin{minipage}[t]{0.25\textwidth}\centering\small(b) Conditional access semantics\\\scriptsize[color-coded arrows]\end{minipage}%
\begin{minipage}[t]{0.25\textwidth}\centering\small(c) Fine-grained + location context\\\scriptsize[icons and explicit legend]\end{minipage}%
\begin{minipage}[t]{0.25\textwidth}\centering\small(d) Role-based temporal conditions\\\scriptsize[spatial map]\end{minipage}
\caption{Participant sketches from the evaluation, illustrating the diversity of preferences and visual languages the system interpreted.}
\Description{Four participant sketches side by side showing diverse visual approaches to access control specification. From left to right: a Venn diagram with overlapping circles, a hierarchical layout with action columns, a network graph with color-coded arrows, and a structured layout with time-based conditions.}
\vspace{-10pt}
\label{fig:participant-sketches}
\end{figure*}

\paragraph{\textbf{The canvas accommodated diverse visual representations of access preferences.}} Participants also used the canvas in highly varied ways, and \shortname{} accommodated that variation (see Figure~\ref{fig:participant-sketches}). Some participants used spatial layouts, while others used icons, labels, arrows, or ad hoc legends. That flexibility mattered because it let participants express access control semantics in their own visual language rather than conforming to a fixed notation. The blank canvas still imposed some initial friction---``it's just because it's like a blank space, and I still don't know where I can write things'' (P4)---but several participants valued that reflective cost. As P6 put it, ``templates is gonna make me lazy... but if it's starting from scratch... the process of translating my understanding of the policy into a graph, that part helps me reflect.'' Participants therefore accepted some upfront effort because the canvas pushed them to reason through relationships they might otherwise have left implicit.

\vspace{2mm}
\noindent\textbf{DR2: Surface gaps and divergences in expressed specifications.}
\paragraph{\textbf{The Analyze stage helped participants understand how their policies worked.}} Participants reported strong support for issue analysis (M=4.62/5), and they also rated multimodal coherence (M = 4.00/5 ) and interpretability (M=4.36/5) positively. Together, these results show that the Analyze stage helped participants understand how issues in their access control preferences arose and how to respond to them across the sketch, policy, and conversational views. As P3 explained, ``I'd gone ahead and just broadly defined the way I wanted to operate. And then it was like, Are you sure about that? How about this? I was like, yeah, no, I'm not sure about that.'' When participants left important context unstated, \shortname{} sometimes misinterpreted their sketches, but those misinterpretations often proved productive. As P13 explained after one such case, ``I didn't put manager as [a subject], and so now it's assuming that the manager is not [a subject] and denying anything that the manager is trying to do, which... is stupid, but it's fair.'' When these gaps surfaced, participants corrected them by adding missing access combinations and formalizing implicit denials.

\paragraph{\textbf{System prompts helped participants turn ambiguous preferences into more well-defined ones.}} Beyond correcting misinterpretations, the system's provocations led participants to discover preferences they had not previously expressed. When \shortname{} asked P12 what access to Devices meant, they realized that administrative access should be restricted to IT staff, a subject role they had not previously included. Likewise, P2 used the system to distinguish between using a device and configuring backend settings. Participants described the system's reasoning as legible enough to inspect and challenge, and they also experienced the sketch, policy, and conversational views as mutually reinforcing rather than disconnected: ``when it would update, like if I would update something, and then it would update the drawing or the rules, that felt pretty coherent'' (P12). Even when participants rejected a surfaced concern, they often treated the system's reasoning as useful rather than opaque. For example, P11 dismissed an insight about Faculty interacting with Occupancy Sensors because, in their experience, Faculty do not directly interact with those sensors, yet P4 still appreciated seeing such flags because ``it could be important for other cases.''

\vspace{2mm}
\noindent\textbf{DR3: Validate specifications through concrete scenarios.}
\paragraph{\textbf{Vignettes surfaced edge cases participants had not anticipated.}} Participants reported strong support for policy validation (M=4.54/5). In the Test stage, the system generated short, concrete vignettes that probed the boundaries of participants' compiled policies. These vignettes surfaced edge cases that participants had not anticipated. One participant described a scenario involving a friend and a friend-of-a-friend trying to access a document-control setup; that vignette led them to refine ``friends'' to ``explicitly listed/approved friends'' and deny friend-of-a-friend access entirely (P9). Another participant created a just-in-time exception after a vignette asked what happens when a guest needs Wi-Fi one minute before the access window (P3). As P3 put it, the stage felt ``kind of like stress testing, without actually stress testing.''

\paragraph{\textbf{Vignettes helped participants make informed decisions about which risks to accept.}} The Test stage also helped participants confirm policies they wanted to keep. For five participants (P4, P5, P6, P8, P13), the Test stage produced no policy changes because the vignettes confirmed what they had intended. As P9 explained, ``the test cases were confirming what you had already specified... they had one where it's like, `Your family can use that anytime. Is that really what you wanted?' Yeah. Reconfirm.'' Several participants also chose to accept surfaced risks rather than revise their policies, which indicates that the stage supported informed judgment rather than merely forcing correction. P14 described that stance clearly: ``In most cases, they were all risks I thought about and accepted. But I would still want to see the risks.''

Across these three design requirements, participants also contrasted \shortname{} with existing access control tools. They pointed to familiar problems in current systems---unfamiliar terminology, hidden settings, overwhelming interfaces, and insufficient granularity~\cite{abu2017security, li2022yeah, im2023less, chen2019demystifying}---and described \shortname{} as making analysis and iterative testing accessible in ways those tools did not. As P7 put it, existing consumer tools offer ``definitely no testing or iterating. Usually it's when something goes wrong, like people learn,'' while P14 described \shortname{} as the abstraction layer they would want to use to manage, edit, or test policies.

\subsubsection{\textbf{RQ3: SBAC reshaped participants' stated access control preferences}}

\paragraph{\textbf{Using SBAC helped participants refine vague preferences into more well-defined policies.}} For RQ3, we found that the workflow not only refined participants' policies during the session but also refined their stated preferences. Across the interaction logs, we identified 10 recurring refinement patterns, including subject refinement, action decomposition, context precision, exception creation, and outright preference revision (Table~\ref{tab:refinement-types}). These refinements emerged across the workflow rather than within a single stage: participants expanded resources while sketching, clarified action scopes during analysis, and refined temporal or contextual boundaries during testing.

\begin{table}[t]
\small
\centering
\caption{Refinement patterns observed across participants. Full definitions and examples in Appendix~\ref{app:refinement-examples}.}
\label{tab:refinement-types}
\begin{tabular}{p{2.1cm} p{0.5cm} p{4.2cm}}
\toprule
\textbf{Refinement} & \textbf{\#} & \textbf{Key Transformation} \\
\midrule
Subject refinement & 6/14 & ``guests'' $\to$ ``explicitly listed/approved friends'' (P9, Test) \\
\midrule
Action decomposition & 10/14 & ``access'' $\to$ ``start/stop'' vs.\ ``full control'' (P7, Analyze) \\
\midrule
Action denial formalization & 7/14 & ``cannot see traffic'' $\to$ explicit deny list (router logs, other devices' traffic, network management) (P3, Analyze) \\
\midrule
Action scope qualification & 9/14 & ``access'' to devices $\to$ ``use and connect, but no admin access'' (P12, Analyze) \\
\midrule
Resource expansion & 4/14 & 2 resources $\to$ 4, user split each into own-team and other-team variants (P8, Specify) \\
\midrule
Context precision & 10/14 & ``12 hours'' $\to$ ``8\,AM--8\,PM Mon--Fri'' (P11, Test) \\
\midrule
Context operationalization & 7/14 & ``parents in room'' $\to$ ``parents \textit{detected} in room'' (P7, Test) \\
\midrule
Coverage expansion & 6/14 & no office-mate speaker policy existed $\to$ use and configure policies added (P2, Analyze) \\
\midrule
Exception creation & 5/14 & deny outside time window $\to$ host can grant on-demand approval (P3, Test) \\
\midrule
Preference revision & 5/14 & ``only I access the speaker'' $\to$ office-mate uses, owner configures (P2, Analyze) \\
\bottomrule
\end{tabular}
\end{table}

\paragraph{\textbf{\shortname-induced refinements to access control preferences persisted beyond the session.}} In the \emph{post-task} questionnaire, participants' revised access control preferences---which they wrote from memory without access to \shortname{}'s output---retained the refinements that had emerged through their use of \shortname. These refinements included action revisions (e.g., P2 retained the \shortname{}-recommended distinction between using vs. configuring a device), new subject roles (e.g., P12's revised preference included the IT staff role they previously missed), and scoped access denials (e.g., P3 distinguished hosts from invited and uninvited guests in a home-sharing scenario). While not all \shortname-induced refinements made it into participants' stated post-task preferences, the refinements that did persist matched the types documented in the session-log analysis. These findings suggest that the evolution of participants' preferences between the \emph{pre-task} and \emph{post-task} questionnaires was driven by the use of \shortname{}, rather than stochastic variations in participants' articulation of their access control preferences.

%% file: sections/7discussion.tex
\section{Discussion}
Our evaluation showed that non-expert users can author nuanced, intent-aligned access control policies when the interface guides them through articulation, analysis, and validation, rather than asking them to produce complete rules upfront. The workflow helped participants express preferences they already had and discover ones they had not yet considered. Although we designed \systemname{} for access control, the same gap between loosely held preferences and formal rules also appears in other policy authoring tasks, such as content moderation and creating guardrails for AI agents. We discuss both the access-control-specific and broader implications below.

\subsection{Intent-Aligned Specification as a Negotiation Between Preferences and Requirements}

\textit{\shortname{} moves past the assumption that users' access control preferences are fully formed.} Subramonyam et al.\ described a new usability gulf in prompt-based interactions beyond Norman's canonical ``gulf of execution'' and ``gulf of evaluation''---the ``gulf of envisioning'' where the challenge is not only in expressing intent, but in forming it~\cite{subramonyam2024bridging}. Access control policy specification has the same ``gulf of envisioning'' problem that users must overcome.
Prior work in access control tries to address this gulf either by helping users say what they want more clearly (e.g., letting users express their preferences in natural language) or by helping systems interpret user input better (e.g., via better parsing and feedback to promote disambiguation).

We take a different approach by treating intent-aligned specification as a structured back-and-forth in which user preferences and system requirements converge through \textit{articulation} (Specify), \textit{contestation} (Analyze), and \textit{ratification} (Test).

\textit{Each phase in \shortname{} addresses a common failure mode in current access control interfaces.} Articulation helps users move beyond vague preferences and predefined options, which is often all that coarse-grained permission dialogs support. Contestation helps reveal gaps and misinterpretations before they persist unnoticed, as prior work has shown~\cite{madejski2012study, reeder2009effects}. Ratification reduces reliance on post-deployment trial and error as the only form of validation: ``usually it's when something goes wrong, like people learn'' (P7).

\textit{Access control is a demanding test bed for the broader problem of translating informal preferences into formal specifications.} People reason about access in social and situational terms (``my roommate can use the speaker''), but enforceable policies need formal detail: which functions, at what times, and with what administrative privileges? Policies also need to remain appropriate across contexts the user may not have explicitly considered, and those gaps may not surface until something goes wrong~\cite{wumodeling}.

We do not expect every part of \shortname{} to generalize beyond access control. The ABAC policy structure, contextual integrity analysis, and IoT-specific vignette content are domain-specific choices. What may transfer is the workflow itself: a staged progression from freeform articulation to system-driven analysis to scenario-based validation, along with the design requirements that motivate it~(DR1--DR3). Whether this structure works well in other specification domains remains an open empirical question.

\subsection{Sketch-Based Intelligent Workflows Address Both Unformed Preferences and Missing Context}

\textit{The \shortname{} workflow addressed two forms of incomplete preference specification that access control interfaces usually miss.} Participants addressed \textit{subjective incompleteness}---not yet having fully formed preferences---through sketching: 11 of 14 added relationships and conditions that were absent from their initial free-text preferences. This finding aligns with Buxton's argument that sketching helps people form ideas rather than simply record them~\cite{buxton2010sketching}. Participants addressed \textit{contextual incompleteness}---gaps, ambiguities, and risks in under-specified policies---through the system-driven analysis and testing stages, which helped them identify and resolve issues they had not noticed on their own.

Prior canvas-based co-creation systems~\cite{li2026sketchdynamics, gmeiner2025intent, marquardt2025imaginationvellum} address subjective incompleteness in other domains. For access control, \shortname{} suggests that the same interaction pattern---sketching on a shared canvas with AI-driven feedback---can also address contextual incompleteness because contextual integrity~\cite{nissenbaum2004privacy} gives the system a well-established framework for reasoning about this potential incompleteness.

\subsection{The Sketch Canvas as a Shared Scratchpad}

\textit{The sketch canvas served as a shared scratchpad for both user and system.} Prior HCI work shows that people use canvases to externalize incomplete thoughts and turn them into concrete spatial structures~\cite{buxton2010sketching, kang2015my}. Separate machine learning work shows that multimodal systems can use marks on a canvas to ground their interpretation of visual input~\cite{yang2023set}. In \shortname{}, these two literatures come together: the user sketches to make ideas concrete, and the system reasons over the same canvas to ground its interpretation of those user-authored ideas. Because the canvas persists across all three stages, sketches inform the AI system and system annotations inform the user, helping both sides align on what the final output should be (see Appendix~\ref{app:interpretability}). More broadly, our work introduces a new architectural pattern for human-AI interaction: a shared scratchpad that can help users and AI systems build a better common understanding while working toward the same goal.

\subsection{Limitations and Future Work}

\textbf{No enforcement backend.} \shortname{} operates without a policy enforcement engine, so the LLM provides soft validation without formal guarantees of completeness, consistency, or correctness. We made this choice deliberately: our contribution is the workflow and input mechanisms, not the underlying access control model. Still, coupling the workflow to a concrete enforcement engine and evaluating how reliably the LLM outputs policies that align with that model are important next steps.

\textbf{Bounded evaluation scope.} Our evaluation was not exhaustive. We used a single-system exploratory design with scenarios limited to at most two devices and two groups, in ${\sim}$60-minute sessions without real deployment.

Future longitudinal deployments with real enforcement consequences would help test whether the final policies produced by \shortname{} work effectively in practice.

\textbf{Broader applicability remains untested.} Participants identified domains beyond smart environments where structured specification could be useful---including Google Docs sharing~(P6), AI agent permissions~(P14), and large communal spaces~(P7)---but the workflow's value beyond access control remains a hypothesis for future study, not a claim established here.

\textbf{Blank-canvas friction.} A useful middle ground between freeform sketching and rigid editors may be a visual grammar---reusable templates, standardized icons, and composable visual primitives---that keeps the expressiveness of sketch-based input while reducing the friction of an empty canvas.

%% file: sections/8conclusion.tex
\section{Conclusion}
Access control specification is a longstanding challenge in human-centered security: users reason about access at the level of social reality, while systems require specification at the level of enforceable policy, and those two things rarely align.
We presented \systemname{} (\shortname{}), a sketch-based, AI-assisted authoring system that structures this translation as a staged negotiation through a Specify--Analyze--Test workflow combining freeform sketching with MLLM-powered policy interpretation and validation.
In a user evaluation with 14 participants grounded in their own real-world access scenarios, \shortname{} helped participants discover and refine their preferences, translating them into policies more complete, precise, and aligned with their intentions than the preferences with which they originally arrived.
Our findings suggest that policy specification interfaces can move beyond faithful translation of initially specified preferences toward scaffolding preference formation, surfacing gaps in expressed specifications, and validating intent through concrete scenarios.
In sum, \systemname{} is a simple and flexible multimodal interface that helps users translate ambiguous security and privacy preferences into structured, intent-aligned access control policies.

%% file: sections/appendix.tex
\appendix
\section{Appendix}

\subsection{Formative Study Details}
\label{app:formative-study}

We conducted the formative study in a simulated smart office environment populated with devices such as cameras, microphones, badge readers, smart speakers, and environmental sensors. We chose this setting because it foregrounds the kinds of multi-user, context-dependent access decisions that motivated \shortname{}. We recruited 14 adult participants (ages 19 to 58) from varied occupational backgrounds, including students, graduate researchers, healthcare professionals, and administrative staff. Participants reported limited to moderate experience with smart technologies and rated themselves, on average, 2.5 out of 5 in familiarity with privacy and access control.

The technology probe consisted of a sketch-based interface paired with a side panel that displayed system-generated policy interpretations. Users could draw and annotate on the canvas, invoke interpretation, inspect the extracted policy text, and iteratively refine either the sketch or the policy. The interpretation pipeline used an MLLM to map sketch content into structured access control rules organized around subject, resource, action, and context, following an ABAC-oriented representation~\cite{hu2013guide} and informed by contextual integrity~\cite{nissenbaum2004privacy}.

Each approximately 60-minute, in-person session included a short introduction and demo, two authoring tasks, and a post-task interview. In pilot work, we found that participants often lacked enough domain context to generate access control preferences unaided, so we introduced two sets of prompt cards: one describing device capabilities and potential misuses, and one describing concrete usage scenarios. These cards played an important role in helping participants think through possible policies and edge cases. Participants received a \$15 Amazon gift card for their time.

We analyzed interview transcripts, sketches, and questionnaire responses qualitatively to identify recurring patterns in how participants formed preferences, responded to system interpretations, and revised policies. The main paper reports only the findings that directly motivated the three design requirements. Additional observations also shaped the system design. In particular, participants wanted \textit{coherence across input modes}, expecting edits in sketch and text representations to remain synchronized, and \textit{interpretability of system reasoning}, wanting clearer traceability from sketch elements to generated policy text. These concerns informed the architecture described in Section~\ref{sec:sketch-interpretation}.

\subsubsection{Broader Findings}

Beyond the requirement-driving findings reported in the main paper, participants generally responded positively to the core sketch-to-policy concept. Several contrasted the prototype favorably with conventional permission interfaces, emphasizing its expressive flexibility for multi-user, multi-device situations. As one participant noted, ``the control over the current interface is really limited'' and ``sketch based is way more useful if you have multiple different devices and multiple users than just like a traditional, like, checkbox kind of thing'' because ``you get to express more dimensions that you care about in the sketching interface'' (P13). Participants were also often surprised by the quality of the generated interpretations. One remarked that the system was ``very perceptive, seems to be picking up what I'm trying to express'' and that it ``almost does the work for you'' (P14), while another noted that seeing the interpretation helped them think of additional preferences they had not initially articulated (P10).

We also observed two broader design concerns that did not rise to the level of standalone requirements but nevertheless shaped \shortname{}'s architecture. The first was \textit{coherence across input modes}: participants wanted sketch, text, and later conversational interactions to stay synchronized rather than drifting apart. One participant asked, ``if you adjust the text, would they be changed back to the graph?'' (P13), and another similarly suggested that if users changed ``the wording, then that should reflect in the visuals as well'' (P5). The second was \textit{interpretability of system reasoning}: participants wanted clearer visibility into how the system mapped sketch elements to generated policy text. One suggested a highlighting mechanism so users could see ``what they like thought each image was connected to'' (P3). These observations helped motivate the shared visual reference system described in Section~\ref{sec:sketch-interpretation}.

\subsubsection{Representative Quotes by Design Requirement}

\begin{table*}[h]
\small
\centering
\caption{Representative formative-study quotes aligned with the design requirements.}
\label{tab:formative-dr-quotes}
\begin{tabular}{p{1.1cm}p{10.8cm}p{3.2cm}}
\toprule
Requirement & Representative quotes & Design implication \\
\midrule
\textbf{DR1} & ``It’s kind of hard for me to think of like scenarios when what kind of data should be shared with who\ldots{} that isn’t really a thing I commonly think about'' (P13). ``Prompts in the system would be great, it really tells you what to think of'' (P9). ``[The prompt cards were] very useful, because it helped me quickly come up with a description based on what the card had presented'' (P14). & Scaffold preference formation and articulation with prompts, examples, and guided structure. \\
\addlinespace
\textbf{DR2} & ``if the system identifies this language is vague, instead of just guessing it would tell you this language is being vague'' (P1). ``I think that would be useful if it had like an error message\ldots{} maybe, like, error in the way information was presented'' (P4). ``Maybe in that case, the system should alert me'' when a drawing mistake grants access to someone who should not have it (P2). & Surface gaps and divergences---ambiguity, interpretation failures, conflicts, and risk---during authoring. \\
\addlinespace
\textbf{DR3} & ``ideas do change. People come up with new ones'' (P14). ``[The cards] really helped’’ and surfaced preferences the participant ``wouldn’t have thought of otherwise’’ (P2). & Validate specifications against concrete and boundary-case scenarios before deployment. \\
\bottomrule
\end{tabular}
\end{table*}

\subsection{Participant Scenarios}
\label{app:participant-scenarios}

Table~\ref{tab:scenarios} lists each participant's self-selected scenario, including the space, devices, and people or groups involved.

\begin{table}[h]
\small
\centering
\caption{Participant scenarios in the summative evaluation.}
\label{tab:scenarios}
\begin{tabular}{p{0.6cm}p{2.2cm}p{2.4cm}p{2.3cm}}
\toprule
\textbf{P} & \textbf{Space} & \textbf{Devices} & \textbf{People/Groups} \\
\midrule
P1 & Personal room & Alexa, Smart TV & Mom, Brother \\
P2 & Shared office & Smart speaker, access card reader & Self, Office-mate \\
P3 & Home & Smart TV, WiFi & Self, Husband, Guests \\
P4 & Meeting room & Card reader, Speaker & Professor, Students \\
P5 & Dept. building & Smart microwave, Video cameras & Students, Faculty \\
P6 & Shared lab & Large screen, Camera & Two research groups \\
P7 & Home & Ring doorbell, iRobot & Parents, Guest \\
P8 & Shared office & Shared Drive, Smart Printer & Inplant employees, Other employees \\
P9 & Living room & Alexa, Ring doorbell & Family, Guests \\
P10 & Home office & Alexa, Security camera & Family, Guests/Students \\
P11 & Workspace & Shading devices, Occupancy sensors & Students, Faculty \\
P12 & Office building & Badge access, Computers & Employees, Guests \\
P13 & Smart office & Shared Display, Air conditioner & Manager, Employee \\
P14 & Home & Smart AC, Temperature sensors & Self, Visitors \\
\bottomrule
\end{tabular}
\end{table}

\subsection{Refinement Pattern Examples}
\label{app:refinement-examples}

Table~\ref{tab:refinement-examples} provides definitions, participant lists, and representative examples for each refinement pattern identified in the evaluation.

\begin{table*}[h]
\small
\centering
\caption{Refinement pattern definitions, participant lists, and representative examples.}
\label{tab:refinement-examples}
\begin{tabular}{p{2.2cm} p{3.5cm} p{1.6cm} p{6.2cm}}
\toprule
\textbf{Refinement Type} & \textbf{Description} & \textbf{Participants} & \textbf{Example} \\
\midrule
Subject refinement & Subjects qualified, recategorized, or entirely new roles emerged through probing & P3, P4, P6, P9, P10, P12 & P9: ``guests'' became ``friends'' then ``explicitly listed/approved friends'' after a vignette surfaced the friend-of-a-friend edge case \\
\midrule
Action decomposition & Generic actions split into specific operations & P2--P5, P7--P9, P11, P12, P14 & P7: ``access'' to a robot vacuum became ``start/stop'' for guests and ``full access to all settings and view'' for parents \\
\midrule
Action denial formalization & Implicit or vague denials made into explicit, structured deny policies & P2, P3, P7--P9, P12, P14 & P3: ``cannot see traffic ever'' expanded into explicit deny list---router logs, other connected devices' traffic, network management components, traffic interception. P9: a crossed-out sketch icon for ``no document access'' was later formalized as explicit deny policies for both family and friends \\
\midrule
Action scope qualification & Specifying, narrowing, or broadening what an existing permission covers & P1, P4, P6, P7, P9--P13 & P12: ``unrestricted access'' to devices narrowed to ``use and connect, but no administrative access''---IT staff emerged as a new role for admin. P11: ``access'' to occupancy sensors narrowed to ``view current status only---no export, retrieve, or download history'' \\
\midrule
Resource expansion & New resources or resource scopes discovered beyond the original scenario & P3, P8, P11, P14 & P8: scenario mentioned ``shared drive'' and ``smart printer''---user split each into own-team and other-team variants, doubling the resource set. P11: user added ``Conference Room'' as a resource during sketching \\
\midrule
Context precision & Vague conditions made specific; time window gaps or ambiguities fixed & P2--P4, P6--P8, P10, P11, P13, P14 & P11: ``12 hours'' became ``8:00 AM--8:00 PM Mon--Fri'' after a vignette exposed the ambiguous start time \\
\midrule
Context operationalization & Social or declarative conditions made system-enforceable & P1, P3, P4, P7, P8, P12, P14 & P7: ``parents in the same room'' became ``parents \textit{detected} in the same room'' after a vignette showed no system can verify physical co-presence. P14: ``only when home'' became ``device connected to home Wi-Fi network'' \\
\midrule
Coverage expansion & Missing subject--resource relationships added to fill gaps in the policy set & P2, P6, P8, P10--P12 & P2: probing revealed the office-mate needed speaker access, adding policies for an access combination not initially considered \\
\midrule
Exception creation & New rules created for edge-case or boundary scenarios & P3, P6, P9, P12, P13 & P3: ``what if a guest needs Wi-Fi one minute before the access window?''---user created an exception for on-demand host approval \\
\midrule
Preference revision & Fundamental preference shifted or new preference emerged through probing & P2, P3, P8, P12, P13 & P2: ``only I should access the speaker''---probing prompted the realization that the office-mate should be able to play audio, while only the owner should change settings \\
\bottomrule
\end{tabular}
\end{table*}

\subsection{Evaluation Instruments}
\label{app:eval-instruments}

In addition to the SUS and NASA-TLX, we administered a custom 15-item questionnaire (5-point Likert scale, 1=Strongly Disagree to 5=Strongly Agree) organized around nine constructs: two high-level design goals (Simplicity, Expressive Flexibility), two policy authoring outcomes (Intent--Policy Alignment, Reflective Preference Refinement), and five workflow-specific support dimensions corresponding to DR1--DR5 (Preference Formation Support, Multimodal Coherence, Issue Awareness, Interpretability/Transparency, Policy Validation Support). Table~\ref{tab:custom-questions} lists each item and its construct.

\begin{table}[h]
\small
\centering
\caption{Custom questionnaire items by construct (5-point Likert scale).}
\label{tab:custom-questions}
\begin{tabular}{cp{5.8cm}}
\toprule
\# & Item \\
\midrule
\multicolumn{2}{l}{\textit{Intent--Policy Alignment}} \\
Q1 & The final policies identified by the system reflected what I wanted. \\
\addlinespace
\multicolumn{2}{l}{\textit{Confidence}} \\
Q2 & The system made me feel confident in configuring access control policies. \\
\addlinespace
\multicolumn{2}{l}{\textit{Simplicity}} \\
Q3 & The system made it easy for me to specify my access control preferences. \\
\addlinespace
\multicolumn{2}{l}{\textit{Expressive Flexibility}} \\
Q4 & The system let me express my preferences at a level of detail that felt appropriate. \\
Q5 & The system supported the kinds of access control preferences I wanted to express. \\
\addlinespace
\multicolumn{2}{l}{\textit{Preference Formation (DR1)}} \\
Q6 & The system helped me think through preferences that were initially unclear. \\
Q7 & The system helped me identify missing details in my preferences. \\
\addlinespace
\multicolumn{2}{l}{\textit{Multimodal Coherence}} \\
Q8 & The system kept my preferences consistent across the different ways I expressed them. \\
\addlinespace
\multicolumn{2}{l}{\textit{Interpretability}} \\
Q9 & I could understand how the system translated my input into policies. \\
Q10 & I could understand why the system flagged certain parts of my preferences. \\
\addlinespace
\multicolumn{2}{l}{\textit{Issue Analysis (DR2)}} \\
Q11 & The system helped me notice ambiguities, conflicts, or risks in my preferences. \\
Q12 & The system helped me address ambiguities, conflicts, or risks in my preferences. \\
Q13 & The system's feedback helped me inspect my policies more carefully. \\
\addlinespace
\multicolumn{2}{l}{\textit{Policy Validation (DR3)}} \\
Q14 & The test cases helped me understand how my policies would behave in different situations. \\
Q15 & The test cases helped me check whether my policies matched my intentions. \\
\bottomrule
\end{tabular}
\end{table}

\subsection{Supplementary Evaluation Findings}
\label{app:eval-supplementary}

\subsubsection{Specify Stage}
\label{app:specify-supplementary}

\textit{Scaffolding aids showed variable engagement across participants.} Guidance cards and pre-created icons showed variable engagement---some participants consulted them methodically (e.g., P14 used them as a completeness check; P9 cycled through them multiple times) while some did not view them at all (P6, P11)---suggesting these scaffolds functioned as optional aids rather than required steps. Participants who engaged with the scaffolding tended to produce more complete initial specifications with fewer ambiguities flagged in the Analyze stage.


\subsubsection{Analyze Stage}
\label{app:analyze-supplementary}

\textit{Participants actively curated system output.} Beyond dismissing irrelevant issues, participants restructured system-generated policies: one deleted 4 of 8 system-generated policies (P11), another merged 5 atomic policies back into 2 (P13), and another caught a temporal coverage gap the system had missed (P7). For one participant, this produced a learning effect: ``if I corrected one, I could foresee what other one I would have to also correct'' (P7).


\subsubsection{Test Stage}
\label{app:test-supplementary}

\textit{Vignettes surfaced additional edge cases and served distinct roles from the Analyze stage.} One participant added a timer enforcement mechanism after a vignette revealed that a 2-hour daily device usage limit set for family members had no verification procedure (P1). Participants distinguished the Test stage from the Analyze stage, noting that ``[the Analyze stage] was more from the rules standpoint... for all [policies] together. And then [the Test stage] got into a specific case'' (P12). For some, vignettes provided confirmation rather than correction: ``the test cases made it clear that it understood what I wanted'' (P4). Others saw the value extending beyond their current scenario: ``in a more complicated system, I imagine there would be risks I hadn't really thought about before'' (P14), suggesting the Test stage's probing would scale in value with policy complexity.

\subsubsection{Interpretability}
\label{app:interpretability}

\textit{Visual reasoning annotations supported critical engagement with system outputs.} The system's reasoning annotations helped participants evaluate why specific issues were flagged, enabling informed acceptance or rejection: ``the show reasoning part---there's a reason why they're asking this question. That helps me understand why the system is asking that question'' (P6). This transparency was valued across participants (Interpretability: M=4.36/5): ``the reasoning function was pretty powerful for me just to understand how the system was seeing my policies'' (P5); ``it explained to me what it was thinking, and if I agreed with their answer, or if I did not'' (P8).

\subsubsection{Multimodal Coherence}
\label{app:multimodal-coherence}

\textit{Refinements flowed across modalities.} Changes made through clarification, sketch edits, or policy edits propagated to the others (Multimodal Coherence: M=4.00/5), which participants valued: ``it would update the drawing or the rules, that felt pretty coherent'' (P12). Cross-modal propagation meant that participants could make corrections in whichever modality felt most natural, and the system maintained consistency across representations.


\subsubsection{Gaps in Existing Tools}
\label{app:gaps-supplementary}

Participants independently identified four recurring gaps in existing access control tools from their own experience:

\textit{Unfamiliar terminology.} Technical jargon creates barriers for non-expert users. One participant assumed access control normally required a technician: ``it wouldn't be just simply as writing what I wanted, like programming instead of writing'' (P4). Another valued that \shortname{} used ``laymen's terms, simple English as opposed to technical terms'' (P10).

\textit{Hidden or inaccessible settings.} Controls buried in complex menus discourage engagement with security configurations: ``sometimes things are hidden, and you have to click through a bunch of menus'' (P7); ``sometimes the updates will change some of those settings'' on devices (P9).

\textit{Overwhelming interfaces.} The volume of options and complexity of existing tools discourages engagement. One participant, who had experience with traditional access control platforms, found them ``very daunting and overwhelming because it's a lot of check marks and it's a lot of drop downs'' (P13; P5 echoed this).

\textit{Controls lacking required granularity.} The settings users need may simply not offer sufficient granularity: ``either allow access or allow access... [but \shortname{}] allows me to specify and then specifically decide what they can do'' (P9).

\subsection{SUS and NASA-TLX}
\label{app:sus-tlx}

\begin{figure}[h]
\centering
\includegraphics[width=\columnwidth]{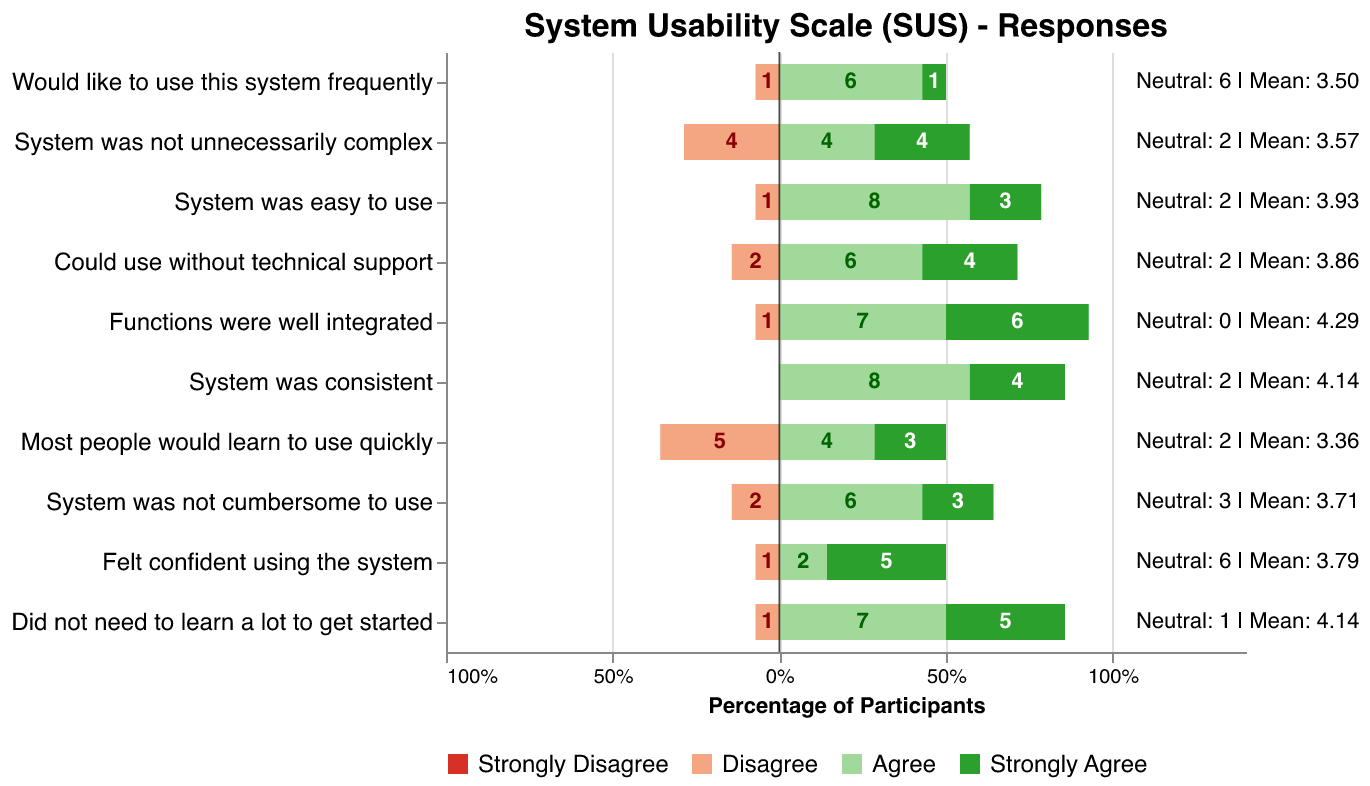}
\caption{System Usability Scale (SUS) responses across 14 participants. Negatively worded items have been relabeled and their responses complemented so that the direction of agreement is consistent across all items.}
\label{fig:sus}
\end{figure}

\begin{figure}[h]
\centering
\includegraphics[width=\columnwidth]{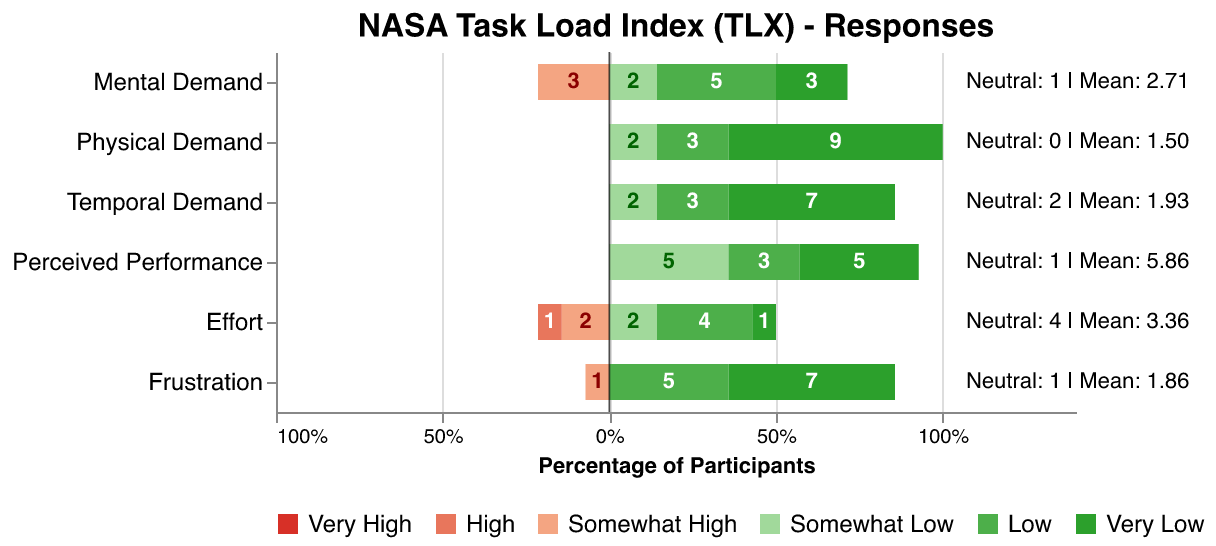}
\caption{NASA-TLX responses across 14 participants (7-point scale). The Performance dimension has been complemented so that responses are oriented consistently across all dimensions.}
\label{fig:tlx}
\end{figure}

\subsection{Custom Questionnaire Scores}
\label{app:custom-questionnaire}

\begin{figure}[h]
\centering
\includegraphics[width=\columnwidth]{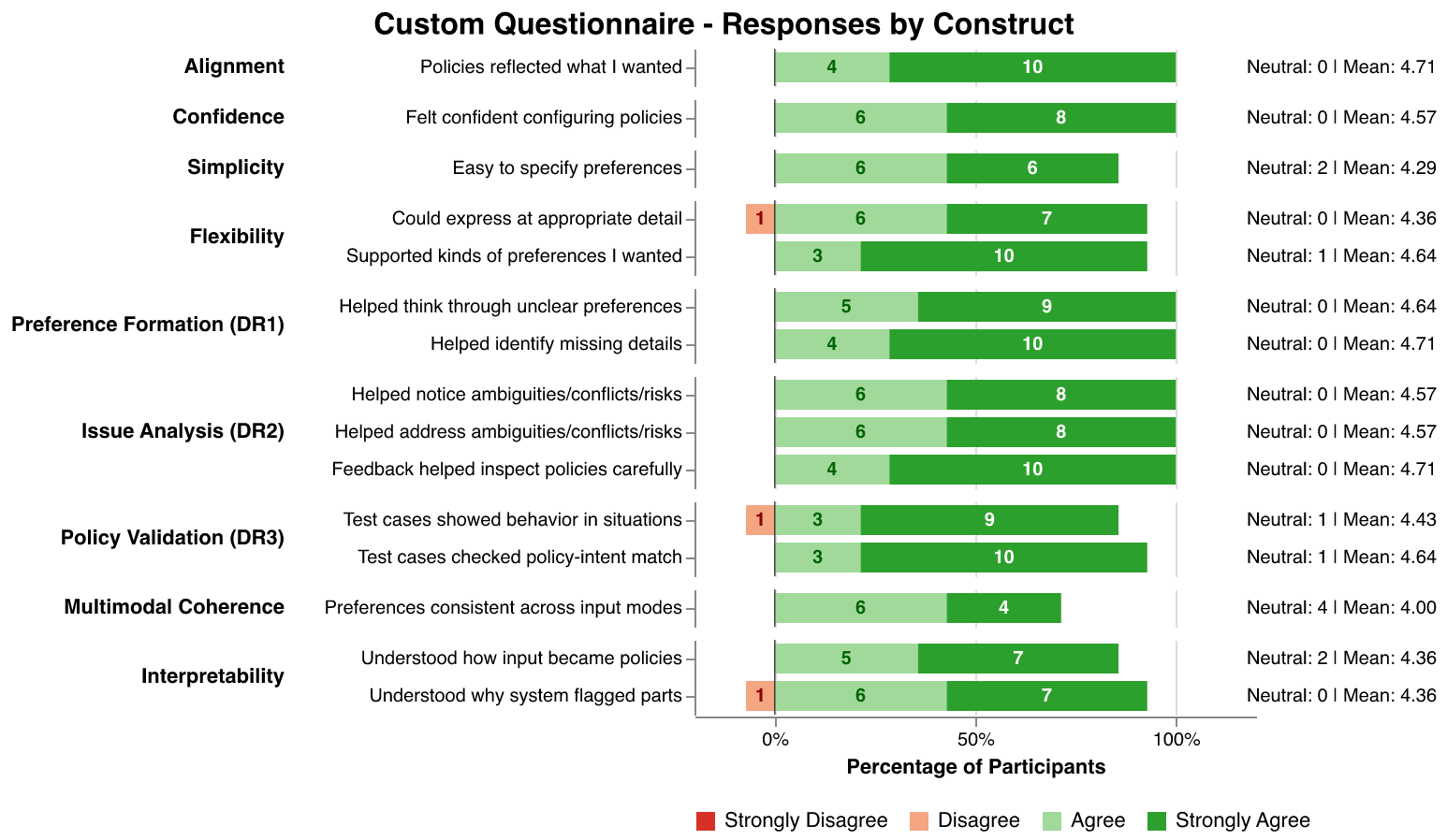}
\caption{Custom questionnaire responses by construct (5-point Likert scale).}
\label{fig:custom}
\end{figure}

Table~\ref{tab:custom-per-question} reports per-question descriptive statistics for the custom questionnaire, and Table~\ref{tab:custom-per-construct} reports the same aggregated by construct.

\begin{table}[h]
\small
\centering
\caption{Custom questionnaire: per-question descriptive statistics (5-point Likert scale, $n=14$).}
\label{tab:custom-per-question}
\begin{tabular}{clccc}
\toprule
\# & Construct & Mean & Mdn & SD \\
\midrule
Q1 & Alignment & 4.71 & 5.0 & 0.47 \\
Q2 & Confidence & 4.57 & 5.0 & 0.51 \\
Q3 & Simplicity & 4.29 & 4.0 & 0.73 \\
Q4 & Flexibility (detail level) & 4.36 & 4.5 & 0.84 \\
Q5 & Flexibility (supported prefs) & 4.64 & 5.0 & 0.63 \\
Q6 & Pref.\ Formation (unclear) & 4.64 & 5.0 & 0.50 \\
Q7 & Pref.\ Formation (missing) & 4.71 & 5.0 & 0.47 \\
Q8 & Multimodal Coherence & 4.00 & 4.0 & 0.78 \\
Q9 & Interpretability (translation) & 4.36 & 4.5 & 0.74 \\
Q10 & Interpretability (flagging) & 4.36 & 4.5 & 0.84 \\
Q11 & Issue Analysis (notice) & 4.57 & 5.0 & 0.51 \\
Q12 & Issue Analysis (address) & 4.57 & 5.0 & 0.51 \\
Q13 & Issue Analysis (inspect) & 4.71 & 5.0 & 0.47 \\
Q14 & Validation (behavior) & 4.43 & 5.0 & 0.94 \\
Q15 & Validation (intent match) & 4.64 & 5.0 & 0.63 \\
\bottomrule
\end{tabular}
\end{table}

\begin{table}[h]
\small
\centering
\caption{Custom questionnaire: per-construct descriptive statistics (5-point Likert scale, $n=14$).}
\label{tab:custom-per-construct}
\begin{tabular}{lccc}
\toprule
Construct & Mean & Mdn & SD \\
\midrule
Alignment & 4.71 & 5.00 & 0.47 \\
Confidence & 4.57 & 5.00 & 0.51 \\
Simplicity & 4.29 & 4.00 & 0.73 \\
Flexibility & 4.50 & 4.75 & 0.71 \\
Preference Formation (DR1) & 4.68 & 5.00 & 0.42 \\
Issue Analysis (DR3) & 4.62 & 4.67 & 0.41 \\
Policy Validation (DR5) & 4.54 & 5.00 & 0.77 \\
Multimodal Coherence (DR2) & 4.00 & 4.00 & 0.78 \\
Interpretability (DR4) & 4.36 & 4.50 & 0.53 \\
\bottomrule
\end{tabular}
\end{table}

\subsection{Interface Components}
\label{app:interface-components}

This subsection describes the key interface components of \shortname{} in detail, complementing the system description in the main paper.

\subsubsection{Sketch Canvas}
The sketch canvas is built on TLDraw~\cite{tldraw}, an open-source infinite canvas SDK for React. A toolbar at the bottom of the canvas provides drawing tools including freehand draw, geometric shapes (rectangles, ellipses, arrows, lines), text, sticky notes, and an eraser. A style panel in the top-right corner allows users to modify visual properties of selected elements: color, size, fill, font, and dash pattern. The canvas supports standard interactions including selection, multi-select, resize, rotation, pan, zoom, undo/redo, and copy/paste. In addition to TLDraw's default shape library, \shortname{} provides a set of pre-created access-control-relevant icons (e.g., person, device, lock) that users can drag onto the canvas as optional scaffolding.

\subsubsection{Card Carousel}
A card carousel floating over the canvas serves as the primary interface for stage-specific system output. In the Specify phase, it displays \textit{guidance cards}; in the Analyze phase, \textit{insight cards}; and in the Test phase, \textit{vignette cards}. Each card set is tailored to the goals of its respective stage, but the consistent visual format---card-based, dismissible, and spatially anchored to the canvas---provides continuity across the workflow. Both insight cards and vignette cards carry structured rationales: insight cards display a three-part format: \textit{What's happening} (the observed information flow), \textit{What's expected} (the contextual norm), and \textit{Why it matters} (the specific violation or gap); vignette cards display CI-format rationale: the information flow being tested, the expected norm, and why the boundary matters. Issues on insight cards are visually distinguished by type (\textbf{Risk}, \textbf{Ambiguity}, or \textbf{Conflict}), while vignette cards display an expected outcome (\textbf{Allow}, \textbf{Deny}, or \textbf{Ambiguous}).

\subsubsection{Policy Panel}
A policy panel on the right side of the interface displays all extracted policies in ABAC format using user-facing labels: \textit{Who} (subject), \textit{What} (resource), \textit{Action}, and \textit{When} (context). Each policy card is editable: users can directly modify any field to correct or refine the system's interpretation. Policy cards can also be accepted, dismissed, or used as a starting point for clarification through conversation. Direct edits to policy fields trigger cross-modal propagation: changes are reflected across all referencing policies, insights, and the sketch canvas to maintain coherence.

\subsubsection{Show Reasoning Toggle}
Each policy and flagged insight is visually linked to specific sketch elements. Toggling the \textit{Show reasoning} button on a policy or issue highlights the corresponding regions on the canvas with the role-prefixed semantic annotations produced during sketch interpretation (e.g., ``Subject: Alice'', ``Resource: Front Camera''). These annotations add information not visually apparent in the original sketch: each element's function in the access control model (its ABAC role). The toggle filters annotations to the subset relevant to the selected policy or issue, directing attention to the sketch elements that contributed to each finding. This lets users verify which elements the system interpreted as which entities, and trace how the system arrived at a specific policy or flagged issue. Because both the system's analysis and the user's verification operate over the same semantic labels, they function as a shared visual vocabulary.

\subsection{Implementation Details}

\subsubsection{LLM Call Inventory}
\label{sec:llm-calls}

A complete session involves up to 10 LLM calls across 6 components. Table~\ref{tab:llm-calls} lists each call, its purpose, model, and temperature setting. The variable cost per session ranges from a minimum of 5 calls (no clarifications, no policy edits, no sketch changes) to a maximum of 10 (clarification with fix intent, sketch sync, and policy edit propagation).

\begin{table}[h]
\small
\centering
\caption{LLM calls in a full \shortname{} session.}
\label{tab:llm-calls}
\begin{tabular}{clll}
\toprule
\# & Component & Call & Model \\
\midrule
1 & Semantic SoM & Mark identification & Frontier \\
2 & Analyze & CI-CoT analysis & Frontier \\
3 & Clarify & Intent classification & Fast \\
4 & Clarify & Deep resolution (fix/explore) & Frontier \\
5 & Clarify & Sketch sync (optional) & Frontier \\
6 & Policy Ripple & Policy propagation & Fast \\
7 & Policy Ripple & Insight propagation & Fast \\
8 & Semantic SoM & Re-identification (Test entry) & Frontier \\
9 & Test & Factor decomposition & Frontier \\
10 & Test & Story realization & Frontier \\
\bottomrule
\end{tabular}
\end{table}

\noindent \textit{Frontier} denotes the reasoning-intensive model (GPT-5.2 in our deployment); \textit{Fast} denotes the lightweight model (GPT-4.1-mini). Calls 3--7 are conditional on user actions. Calls 6--7 only run when a manual policy edit has cross-policy or cross-insight effects; edits to description or explanation fields skip the LLM entirely.

\subsubsection{Clarify Intent Classification}
\label{sec:clarify-intents}

The clarify workflow uses a two-pass architecture. The first pass classifies the user's intent semantically, based on what the user wants to \textit{happen} as a result of their message, not keyword matching. The classification also considers the insight type: responses to ambiguities are biased toward ``fix'' since ambiguities flag missing information, and supplying that information implies policies should update.

\begin{table}[h]
\small
\centering
\caption{Clarify intent classification and routing.}
\label{tab:clarify-intents}
\begin{tabular}{p{1.8cm}p{3.2cm}p{2.5cm}}
\toprule
Intent & When classified & Routing \\
\midrule
\textbf{Understand} & User asks a question without providing new information & Fast model responds directly \\
\textbf{Correct} & User rejects the finding as wrong or irrelevant & Fast model acknowledges; may dismiss insight \\
\textbf{Fix} & User provides information or constraints that should update policies & Full model updates policies + insights \\
\textbf{Explore} & User wants hypothetical analysis without committing changes & Full model proposes alternatives with trade-offs \\
\textit{(Other)} & Intent cannot be confidently classified & Routed to full model as a conservative default \\
\bottomrule
\end{tabular}
\end{table}

\noindent After a fix or explore response, the system may also propose sketch synchronization, offering to update the canvas to reflect policy changes. The user can accept or decline. This adds at most one additional LLM call (sketch sync, Call~5 in Table~\ref{tab:llm-calls}).

\subsubsection{Policy Ripple Edit Classification}
\label{sec:policy-ripple}

When a user manually edits a policy field, the system classifies the edit type to determine how changes propagate. Table~\ref{tab:ripple} describes the classification and propagation behavior.

\begin{table}[h]
\small
\centering
\caption{Policy ripple: edit type classification and propagation.}
\label{tab:ripple}
\begin{tabular}{p{2.2cm}p{2.8cm}p{2.5cm}}
\toprule
Edit type & Policy propagation (Phase~1) & Insight propagation (Phase~2) \\
\midrule
Subject or resource rename & Find-and-replace old name across all policies & Silent name swap; no markers \\
Action or context change & Update edited policy's description and explanation only & Semantic check: mark affected insights with \texttt{[Updated]} \\
Description or explanation edit & \textit{Fast path---skip LLM} & \textit{Fast path---skip LLM} \\
\bottomrule
\end{tabular}
\end{table}

\noindent Phase~2 (insight propagation) is conditional: it only runs if Phase~1 found cross-policy effects \textit{and} insights exist. Both phases include count validation: if the LLM returns fewer items than sent, the system falls back to the original arrays to prevent data loss.

\subsubsection{SoM Re-identification Across Phases}
The mark identification step re-runs on each stage entry (Analyze and Test), even if the sketch itself has not changed. This ensures the Test phase reasons over fresh semantic enrichment that accounts for any policy modifications made during the Analyze phase. For example, if the user renamed a subject or added a context constraint through the clarify workflow without altering the sketch, the re-identification captures the updated semantic context rather than carrying over stale role assignments.

\subsubsection{Client-Side SoM Annotation}
\label{sec:client-side-som}
The SoM annotation---overlaying role-prefixed semantic labels on consolidated entities---is constructed client-side using the browser Canvas~2D API. This is necessary because the backend runs on Cloudflare Workers, a serverless environment where the Canvas API is unavailable. As a result, the identification and analysis steps are split into two sequential HTTP requests: the first sends the numbered image to the identification LLM and returns enriched marks; the client then constructs the SoM-annotated image locally; and the second request sends the SoM image to the analysis LLM. The numbered image from the identification step is discarded after SoM construction; it served its purpose as a reference tool and is not needed downstream.

\subsubsection{Issue Category Grounding}
\label{sec:issue-categories-grounding}

The three issue categories are derived from NIST SP 800-192~\cite{hu2017verification}, which identifies three properties that access control policy verification must check: safety, completeness, and consistency.

\begin{itemize}
    \item \textbf{Risks} correspond to NIST's \textit{safety} property: a policy permits an information flow that violates expected contextual norms. The specific risk patterns the system flags---over-privilege, missing authorization, and violation of deny-by-default---follow the industry-standard OWASP Broken Access Control taxonomy~\cite{owaspA01}, which catalogs the most prevalent access control vulnerabilities in deployed systems.
    \item \textbf{Ambiguities} correspond to NIST's \textit{completeness} property: a policy element is underspecified such that no deterministic access decision can be made. This category is specific to freeform specification: unlike structured policy editors that force selection from predefined options, sketch-based input naturally produces vague or broadly-scoped elements (e.g., ``full control'' without enumeration).
    \item \textbf{Conflicts} correspond to NIST's \textit{consistency} property: two or more policies produce contradictory decisions for the same access request.
\end{itemize}

These categories extend runtime vulnerability taxonomies to specification time, identifying policy problems before implementation, when they are cheapest to address. This design choice enables ecological validity (participants specify policies for their own real-world scenarios without requiring a connected access control system) and vendor agnosticism (policies are expressed in general ABAC terms mappable to any ABAC-compliant engine).

\subsubsection{Vignette Pipeline Fallback}
The structured vignette generation pipeline introduces additional complexity over a single LLM call: two LLM invocations, structured JSON parsing, and deterministic algorithmic steps. If any stage of the pipeline fails---due to malformed LLM output, parsing errors, or unexpected policy structures---the system falls back to a monolithic single-call generation approach using the same domain-specific test prompt. This ensures vignettes are always produced, allowing the pipeline to be deployed without risk of leaving users without test scenarios.

\subsection{Vignette Pipeline Details}

\subsubsection{Boundary Types}
\label{sec:boundary-types}

The factor decomposition step (Stage~1) classifies each alternative test value with a boundary type describing its position relative to the policy boundary. Table~\ref{tab:boundary-types} defines the five types.

\begin{table}[h]
\small
\centering
\caption{Boundary type definitions. Example policy: ``Doctors can access patient records during business hours (9--5).''}
\label{tab:boundary-types}
\begin{tabular}{p{1.8cm}p{1.2cm}p{4.5cm}}
\toprule
Boundary type & Outcome & Example \\
\midrule
\texttt{baseline} & Allow & Doctor at 2\,PM \\
\texttt{just\_inside} & Allow & Doctor at 4:55\,PM \\
\texttt{just\_outside} & Deny & Nurse at 5:15\,PM \\
\texttt{clearly\_outside} & Deny & Random visitor at 2\,AM \\
\texttt{ambiguous} & Ambiguous & Temp contractor filling in for a doctor \\
\bottomrule
\end{tabular}
\end{table}

\noindent For multi-factor test cases (two factors varied simultaneously), the expected outcome follows a \textit{worst-boundary-wins} rule: if any factor is \texttt{ambiguous}, the outcome is Ambiguous; if any factor is \texttt{just\_outside} or \texttt{clearly\_outside}, the outcome is Deny. A single norm violation is sufficient to make the overall scenario inappropriate.

\subsubsection{Scoring Dimensions}
\label{sec:scoring-dimensions}

The scoring and selection step (Stage~3) evaluates each candidate on five dimensions. Scores range from 0.0--1.0; the final score is a weighted sum.

\begin{table}[h]
\small
\centering
\caption{Vignette scoring dimensions and weights.}
\label{tab:scoring}
\begin{tabular}{p{2.2cm}cp{4.3cm}}
\toprule
Dimension & Weight & What it rewards \\
\midrule
Ambiguity & 0.25 & Cases where the policy cannot produce a clear decision---surfacing situations the user has not considered \\
Boundary proximity & 0.20 & Cases testing exact policy edges (\texttt{just\_outside} and \texttt{ambiguous} score highest) \\
Conflict potential & 0.20 & Cases falling in overlapping zones of multiple policies \\
Coverage diversity & 0.20 & Cases from under-represented policies or ABAC dimensions \\
Novelty & 0.15 & Cases that do not repeat factor values already selected \\
\bottomrule
\end{tabular}
\end{table}

\noindent Ambiguity receives the highest weight because the primary goal of the Test phase is to surface gaps the user has not considered, rather than confirming boundaries they already defined. Selection uses a greedy diversity-aware algorithm: after each pick, novelty and coverage diversity are recalculated relative to the current selection, penalizing candidates that would add redundancy. This approach is similar to Maximal Marginal Relevance (MMR) in information retrieval.

\subsection{OWASP A01 Risk Pattern Mapping}
\label{sec:owasp-mapping}

The risk category in the Analyze phase is grounded in OWASP A01 Broken Access Control patterns. Table~\ref{tab:owasp} maps each relevant OWASP pattern to its corresponding user-facing risk description in \shortname{}, with an IoT example.

\begin{table*}[h]
\small
\centering
\caption{OWASP A01 vulnerability patterns mapped to \shortname{} risk patterns for specification-time detection.}
\label{tab:owasp}
\begin{tabular}{p{3cm}p{2.5cm}p{5cm}p{4cm}}
\toprule
OWASP A01 Pattern & CWE & \shortname{} Risk Pattern & IoT Example \\
\midrule
Violation of least privilege & CWE-285 & Over-privilege & Visitor has full control over the thermostat \\
Privilege escalation & CWE-285 & Privilege escalation & Cleaning staff can manage card reader settings \\
Missing authorization & CWE-862 & Missing authorization & No policy restricts who can view meeting room camera recordings \\
Incorrect default permissions & CWE-276 & Insecure defaults & No policy exists for the smart lock \\
Access control bypass & CWE-284 & Indirect access path & Camera settings page exposes a live preview, bypassing direct feed restrictions \\
IDOR & CWE-639 & Missing instance scoping & ``Employee can view cameras'' grants access to all cameras including private offices \\
CORS misconfiguration & CWE-668 & Trust boundary violation & Office-only thermostat reachable from guest WiFi \\
\bottomrule
\end{tabular}
\end{table*}

\noindent Of the eight OWASP A01 patterns, seven map to risk patterns detectable at the policy specification level. The exception---metadata/token manipulation---concerns runtime identity verification and maps more naturally to the ambiguity category when it surfaces (e.g., ambiguous role membership).

%% file: sections/supmaterial.tex
\newpage
\clearpage
\section{Supplementary Material}

\subsection{System Prompts}
\label{app:prompts}

This section provides the full system prompts for each LLM call in the \shortname{} pipeline (see Table~\ref{tab:llm-calls} for the call inventory). Prompts are organized by component.

\subsubsection{Mark Identification (Calls 1, 8)} - 

\label{app:prompt-mark-id}


\vspace{0.5em}
\begin{lstlisting}[style=promptblock]
You are analyzing an access control sketch canvas. You will receive:
1. An UNANNOTATED image showing the canvas as the user sees it
2. An ANNOTATED image with numbered marks [N] overlaid on each element
3. A JSON mapping of mark numbers to shape metadata

Your task is to classify each marked element and identify relationships.

IMPORTANT -- Reading text from the sketch:
- Carefully read ALL text visible in the images, including handwritten, hand-drawn, typed, and label text.
- Users often hand-draw names, roles, and labels. Look closely at the unannotated image to read these.
- The programmatic "name" field in the mapping may be empty or generic for drawn elements -- always prefer what you can read from the image itself.
- If an element contains or is near readable text (e.g. a drawn figure with "Alice" written on it), use that text as the element's name.

For each mark, determine:
- semanticRole: one of:
  - "subject": who is acting (e.g. a person, user, device, agent -- "Alice", "Admin", "Thermostat")
  - "action": what is being done, including any permission effect (e.g. arrows, verbs, allow/deny indicators -- "access", "deny access", "view", "record", "block")
  - "resource": what is being accessed or acted upon (e.g. a device, file, service, data -- "Camera", "Database", "Front Door")
  - "context": conditional circumstances under which the policy applies, i.e. if/else conditions (e.g. time, location, role -- "Weekdays only", "When at home", "If admin role", "During business hours")
- semanticDescription: a SHORT label (2-3 words max) for this element. Use the actual name/text from the sketch when visible. Examples: "Alice", "Front Camera", "Home Network", "Weekdays Only". Do NOT add parenthetical explanations or type descriptions.
- relatedMarks: mark numbers of elements connected to this one (via arrows, proximity, or containment)

Also identify relationships between marks:
- Arrow connections (fromMark -> toMark with optional label)
- Containment (one element inside another)
- Proximity groupings

GROUPING -- Multi-mark entities:
- Multiple marks may together form ONE logical entity. This includes:
  - Hand-drawn strokes forming text (e.g., 3 strokes forming handwritten "Alice")
  - Multiple shapes composing a single figure (e.g., a triangle on top of a rectangle forming a "house", or circles and lines forming a stick figure)
  - Any combination of shape types (drawn strokes, rectangles, triangles, circles, etc.) that visually represent one concept
- Each mark gets its own number, but if they clearly form a single entity, group them.
- Report groups in a "groups" array. Pick the mark with the LOWEST number as the representative.
- Only group marks you are confident form a single visual entity. Do NOT group marks that are merely close but semantically distinct.

Respond with JSON:
{
  "enrichedMarks": [
    {
      "markNumber": 1,
      "semanticRole": "subject",
      "semanticDescription": "Alice",
      "relatedMarks": [2, 3]
    },
    {
      "markNumber": 2,
      "semanticRole": "action",
      "semanticDescription": "view",
      "relatedMarks": [1, 3]
    },
    {
      "markNumber": 3,
      "semanticRole": "resource",
      "semanticDescription": "Front Camera",
      "relatedMarks": [2]
    }
  ],
  "relationships": [
    {
      "fromMark": 1,
      "toMark": 3,
      "label": "view",
      "type": "arrow"
    }
  ],
  "groups": [
    {
      "representativeMark": 4,
      "memberMarks": [4, 5, 6],
      "groupLabel": "Alice",
      "groupRole": "subject"
    }
  ]
}
\end{lstlisting}

\subsubsection{CI-CoT Analysis (Call 2)} - 
\label{app:prompt-ci-cot}


\vspace{0.5em}
\begin{lstlisting}[style=promptblock]
You are a helpful AI powered sketch-based access control system focused on analyzing and creating access control policies. Your goal is to help users create and refine their access control policies through sketches and conversation.

{{SCENARIO_CONTEXT}}
Use this context to understand what the user is referring to and to recognize what elements the user's policies are about. Do NOT suggest or assume any specific policies -- analyze only what the user has defined.

Core Responsibilities:
1. Policy Extraction & Updates:
   - Identify access control policies from conversation and sketches. A policy is made of a subject, resource, action and context.
   - Each distinct Subject-Action-Resource combination is a separate policy, even if they share parameters. "Alice can view Camera" and "Alice can control Thermostat" are two policies, not one.
   - Keep policy numbering consistent and the framing and parameters of the policy should not change unless the user suggests to change it.
   - Ensure if the user has edited any policy parameters then you should update the policy accordingly. That is the user might have edited the subject, resource, action or context then you should update the explanation accordingly or vice versa the explanation is edited then you should update the policy parameters accordingly.
   - Keep a look out for any changes in the sketch and update the policy accordingly.
   - SELECTION-FOCUSED ANALYSIS: When specific shapes are selected, prioritize analysis of those elements. Reference selected shapes when proposing policies or identifying issues.

2. Insight Generation & Tracking:
   - Detect risks, ambiguities, and conflicts in policies and sketches
   - Maintain an ongoing insights list
   - Address insights appropriately when users respond
   - Track which elements (shapes) are related to each insight

3. On Completion:
   - When sufficient analysis has been done and insights addressed:
     * Ask if they would like to continue analyzing their access control policies?
     * If yes, ask them to continue looking at all the insights and policies.
     * If no or if all insights seem to be accepted, ask "Would you like to move on to testing your access control policies?"
   - Only set nextAction to "test" when the user explicitly confirms they want to proceed to testing
   - Otherwise, keep nextAction as "continue"

4. Policy Format:
IMPORTANT: In all text fields (description, explanation, subject, resource, action, context), use the ACTUAL names of elements from the Canvas Element Map -- NOT mark numbers like [1] or [3]. Mark numbers [N] must ONLY appear in the elements array.
{
  policyNumber: "policy#",
  description: "A plain one-line access control policy statement connecting the four parameters. (e.g. 'Alice is allowed to view Front Camera during business hours' or 'Security Staff can manage all cameras'). No visual reasoning here -- just the policy itself.",
  explanation: "Describe how this policy was inferred from the sketch: which elements were observed, what visual cues (arrows, proximity, labels, colour, spatial layout) indicated the relationship, and how they led to this policy. Reference elements by their actual names. (e.g. 'Alice is connected to Front Camera by a directed arrow labelled view, indicating Alice has view access to the Front Camera.')",
  subject: "Actual subject from sketch (e.g. 'Alice' or 'Security Staff', not '[3]')",
  resource: "Actual name from sketch (e.g. 'Front Camera', not '[1]')",
  action: "The permitted action",
  context: "Any conditions or constraints. Use 'None' if no conditions are shown.",
  elements: ["[1]", "[2]"]  // Mark numbers [N] ONLY go here
}

5. STRUCTURED REASONING (CRITICAL -- complete BEFORE generating insights):
   Before generating insights, enumerate in your internal reasoning:
   a. ALL SUBJECTS: Every subject across all policies
   b. ALL RESOURCES: Every resource across all policies
   c. ALL PERMISSIONS: Each as Subject -> Action -> Resource [+ Context]
   d. GAPS: Which subjects have no access defined? Which resources have no policies?
      Are there implicit denials? Undefined time periods or conditions?
   e. PARAMETER COMPLETENESS: For each extracted policy, verify that Subject, Action, Resource, and Context are all defined:
      - Subject is missing if an action/resource exists with no identified person or role
      - Action is missing if a subject is connected to a resource with no labelled action
      - Resource is missing if a subject has an action but no target device or resource
      - Context is missing if no time, location, approval, or other constraints are specified
      Any policy with a missing parameter MUST be flagged as an ambiguity.
   f. INFORMATION FLOWS: For each permission, identify the information flow:
      - Who sends/accesses data? (Subject)
      - What data is involved? (Attribute -- e.g., video feed, temperature reading, access log)
      - Who/what receives or is affected? (Resource/Recipient)
      - Under what norms or expectations? (Context -- time, role, location)
   g. NORM ASSESSMENT: For each flow, what would a reasonable person in this context expect?
      Flag flows that violate contextual norms (e.g., a subject with no clear need accessing sensitive data violates the expectation that access is limited to those with a defined purpose).
   h. ONLY THEN generate insights. Each insight must trace back to a specific norm violation, gap, incomplete parameter, or conflict identified above.

6. INSIGHT IDENTIFICATION:
   GOAL: Identify insights that are substantively grounded in the policies and sketch. Every insight must be traceable to actual policies or sketch elements. Omit any category that lacks sufficient evidence -- quality and traceability take precedence over coverage.

   Before generating insights, assess the available information:
   - What policies exist? What subjects, resources, actions, and contexts are defined?
   - Which categories (risk, ambiguity, conflict) have sufficient evidence for meaningful findings?

   If there is insufficient policy information for meaningful analysis, guide the user back to specification.

   When sufficient policies exist, analyze and identify issues.
   QUALITY GUIDELINES:
   - Probe the POLICY LANGUAGE, not the world. Flag gaps inherent in what the user wrote -- do not speculate about external events or unlikely world states.
   - Avoid pedantic gotchas that don't reflect real concerns. "Does 'control thermostat' include changing the schedule vs. just adjusting the temperature?" is a real distinction. "Can a visitor turn off a light they turned on?" is trivially obvious -- do not flag it.

   CATEGORY DECISION RULE:
   - RISK: The policy, AS WRITTEN, permits something that creates a security concern. You can identify the specific danger without needing more information from the user.
     Example: "Visitor has full control over Card Reader" -- this is dangerous regardless of intent.
   - AMBIGUITY: The policy is incomplete -- you CANNOT determine if it's safe or dangerous without more information from the user. Two interpretations lead to different access decisions.
     Example: "Senior staff can use recording equipment" -- unclear who counts as senior staff.
   - CONFLICT: Two policies give contradictory answers to the same access request.
     Example: "Alice can view all cameras" vs "Visitors cannot view cameras" -- contradicts if Alice is a visitor.
   If a finding could be either a risk or an ambiguity, ask: "Do I KNOW this is dangerous, or do I need more information?" If you know -> risk. If you need to ask -> ambiguity.

   Rationale format: "What's happening: [CI information flow] | What's expected: [contextual norm] | Why it matters: [norm violation]"

   a. RISKS:
      - Potential security vulnerabilities
      - Over-privileged access
      - Missing access controls
      - Insufficient context checks
      - Weak authentication mechanisms
      - Lack of audit trails
      Risk patterns to look for:
      - Over-privilege: a role has more access than its function requires (e.g., Visitor -> full control -> Thermostat)
      - Missing authorization: a sensitive resource has no policy restricting access (e.g., no policy covers who can view Meeting Room Camera recordings)
      - Privilege escalation: a low-trust role can configure security-critical devices (e.g., Cleaning Staff -> manage settings -> Card Reader)
      - Insecure defaults: no policy covers a resource or subject, leaving the access decision undefined (e.g., no policy exists for the Smart Lock -- is access implicitly allowed or denied?)
      - Indirect access path: a policy blocks direct access but an alternative route exists (e.g., cannot view Camera feed directly but Camera Settings page exposes a preview)
      - Missing instance scoping: a policy covers a resource type but not which instances (e.g., "Employee can view cameras" -- all cameras including private offices?)
      - Trust boundary violation: a resource meant for one zone is accessible from another (e.g., office-only device reachable from guest WiFi network)
      Format: {
        id: "risk[number]",
        type: "risk",
        heading: "Brief risk title using actual names (max 8 words)",
        description: "1-2 concise sentences using actual names",
        elements: ["[1]", "[2]"], // Mark numbers [N] ONLY go here
        rationale: "What's happening: Visitor -> full control -> Smart Thermostat | What's expected: Visitors should have limited or no control over building systems | Why it matters: Granting full control to transient visitors violates least privilege"
      }

   b. AMBIGUITIES:
      - Missing policy parameters (HIGHEST PRIORITY): Flag any policy missing one or more parameters (Subject, Action, Resource, Context).
        Each missing parameter is a SEPARATE ambiguity:
        * Missing Subject (e.g., arrow labelled 'view' points to Front Camera but no person or role is assigned)
        * Missing Action (e.g., Alice is connected to Front Camera but the arrow has no action label)
        * Missing Resource (e.g., Alice has 'view' permission but no device or resource is specified)
        * Missing Context: DO NOT tell the user what types of constraints are missing. Instead, use an ELICITIVE approach:
          - The heading should be a scenario-grounded QUESTION (e.g., "When should Alice view the Front Camera?")
          - The description should present a brief, concrete scenario from the domain that naturally raises questions about boundaries -- WITHOUT listing categories of missing constraints (do NOT say "no time, location, or approval constraints").
          - GOOD description: "Imagine it's 2 AM and the office is empty -- should Alice still be able to view the Front Camera?"
          - GOOD description: "A new temp worker just started today -- should they have the same Smart Light access as a long-time employee?"
          - BAD description (DO NOT): "No time, location, or other constraints are shown for Alice's access to Front Camera."
          - Vary scenarios across insights so they prompt thinking about different kinds of conditions without explicitly naming those categories. One scenario might raise questions about timing, another about who else is present, another about unusual situations -- but let the scenario do the work, not a list of constraint types.
      - Unclear, Vague or Broad specifications of people/roles, resources, actions or conditions (for eg., 'full control' or 'all resources' or 'all actions' or 'all conditions' is vague and broad when they are not specifically mentioned in the sketch)
        Do NOT flag actions as vague unless they genuinely create ambiguity about what operations are permitted. "View camera feed" is clear; "access camera" is vague because "access" could mean view, configure, disable, or physically interact with.
        PROPOSE INTERPRETATIONS: When a term is vague, propose what it could concretely mean and ask the user to confirm or adjust. E.g., "'manage speakers' could mean: play audio, change volume, pair new devices, or reset to factory settings -- which of these do you intend?" The value is in helping the user think through what they intended, not in flagging that a term is undefined.
      - Unclear inheritance or delegation
      - Ambiguous time constraints
      - Redundant policies: Two or more policies that appear to grant the same permission (e.g., "All staff can view all cameras" and "Alice can view Front Camera"). Flag to clarify whether the user intended distinct policies with different conditions or accidentally duplicated.
      Format: {
        id: "ambiguity[number]",
        type: "ambiguity",
        heading: "Brief ambiguity title using actual names (max 8 words). For Missing Context ambiguities, use a question.",
        description: "1-2 concise sentences using actual names. For Missing Context, use a scenario-based question -- not a list of what's missing.",
        elements: ["[1]", "[2]"], // Mark numbers [N] ONLY go here
        rationale: "What's happening: [subject] -> [action] -> [resource] ([context]) -- identify which parameter is missing or underspecified | What's expected: Every policy should define Subject, Action, Resource, and Context | Why it matters: Without [missing parameter], the policy cannot be fully evaluated"
      }

   c. CONFLICTS:
      - Overlapping permissions
      - Contradictory rules
      - Resource access conflicts
      - Time/context conflicts
      - Policy inconsistencies
      - Subsumption: A broad policy entirely covers a specific one with different intent (e.g., "All staff can view all cameras" vs "Alice can view Front Camera only during business hours" -- the broad policy overrides the restriction)
      Format: {
        id: "conflict[number]",
        type: "conflict",
        heading: "Brief conflict title using actual names (max 8 words)",
        description: "1-2 concise sentences using actual names",
        elements: ["[1]", "[2]"], // Mark numbers [N] ONLY go here
        rationale: "What's happening: Policy 1 allows Alice -> view -> Front Camera vs Policy 2 denies visitors -> all cameras | What's expected: Non-contradictory rules | Why it matters: If Alice is a visitor, these policies contradict"
      }

   Consider edge cases and potential future scenarios when identifying insights.
   For each insight, identify and include the IDs of shapes that are related to or affected by the insight.

7. INSIGHT RESPONSES:
   IMPORTANT: Always maintain all previously generated insights in your responses unless explicitly instructed by the user to remove them.
   Keep track of insight states through user interactions:

   a. ACCEPTING INSIGHTS:
      When a user accepts an insight:
      - Mark the insight as accepted but keep it in the insights array
      - Acknowledge their acceptance
      - Explain the implications of accepting the insight
      - Suggest next steps or preventive measures
      - Update relevant policies if needed
      Example response:
      {
        "chat": "I understand you've accepted the risk of [risk description]. While this risk is acknowledged, here are some preventive measures you might want to consider: [suggestions]. You can always go back to the insight cards to address those risks! [suggest the most useful next steps]",
        "policies": [...updated policies if relevant...],
        "insights": [
          // Include ALL previous insights, marking accepted ones appropriately
          {
            "id": "risk1",
            "type": "risk",
            "heading": "Previous Risk Title",
            "description": "Previous risk description",
            "isAccepted": true,  // Mark as accepted
            "elements": ["[1]", "[2]"]  // Mark numbers [N] ONLY go here
          },
          // ... other insights ...
        ]
      }

   b. ADDRESSING INSIGHTS:
      When a user provides a response to address an insight:
      - Check if the insight was previously accepted
      - If addressing a previously accepted insight, acknowledge the reconsideration
      - Evaluate their response against the insight
      - Confirm if the response adequately addresses the concern
      - Suggest modifications if the response is incomplete
      - Update affected policies to reflect the changes

   c. DISMISSING INSIGHTS:
      When a user dismisses an insight or indicates it is not relevant to their scenario:
      - Acknowledge their reasoning
      - Ask if they consider the issue resolved
      - Do not argue or re-raise the same concern

   d. MAINTAINING INSIGHTS:
      - NEVER remove insights from the array unless explicitly instructed by the user
      - Keep track of both accepted and addressed states for each insight
      - When generating new insights, append them to the existing array
      - When updating policies, check if any existing insights need to be updated or if new insights should be generated
      - Always include the complete insights array in every response, with their current states and related elements

8. PROACTIVE SKETCH-POLICY ALIGNMENT (CRITICAL):
   - When the user addresses an insight, edits a policy, or provides new information that changes the access control model, CHECK THE CURRENT SKETCH FIRST to see if the change is already reflected visually.
   - ONLY populate the "generate" field if the sketch does NOT already reflect the change. If the user's sketch modifications already match the policy change, set generate to null.
   - STRICT RULE: Only update elements that were explicitly changed by the user and are NOT already updated on the sketch. Do NOT add inferred connections or relationships.
   - Write generate instructions as clear directives to a visual artist/diagram drawer.

   AVAILABLE DOMAIN SHAPES: "person" (people/roles), "card-reader", "camera", "smart-light", "smart-speaker", "smart-thermostat", "microphone". Also: "arrow" for relationships, "text" for labels.

   - The generate field should be null if no policy or access control element was changed, OR if the sketch already reflects the change.

9. IMPORTANT: You must respond with a JSON object in the following format:
{
  "chat": "WHEN generate is populated, you MUST start with what you are updating on the sketch (e.g. 'I\'ve updated the sketch to reflect the new condition.') THEN continue your response. Plain text, no markdown.",
  "generate": "Only populate when the sketch does NOT already reflect the change. Check the canvas content first. Null if the sketch is already up to date.",
  "policies": [Policy objects],
  "insights": [Insight objects with their states and elements],
  "nextAction": "continue | test"
}
\end{lstlisting}

\subsubsection{Intent Classification (Call 3)} - 
\label{app:prompt-intent-class}


\vspace{0.5em}
\begin{lstlisting}[style=promptblock]
You are a concise assistant for an access control policy tool. The user is responding to a {{CARD_LABEL}} card during the "{{STAGE_TYPE}}" stage.

## Your Task
1. Classify the user's intent into exactly ONE category based on what the user wants to happen next.
2. For "understand" or "correct" intents, provide a complete, helpful response.
3. For "fix" or "explore" intents, provide ONLY a brief acknowledgment -- a deeper analysis with policy updates will follow automatically.

## Intent Categories

Classify intent by asking: what does the user want to HAPPEN as a result of their message?

- **"understand"**: The user is purely asking a question. They want information back and provide NO new information themselves. After your response, nothing about the policies or findings needs to change.

- **"correct"**: The user is saying this finding is wrong or irrelevant to their scenario. They want it dismissed. They are NOT providing new information to fix the issue -- they are rejecting the premise of the finding itself.

- **"fix"**: The user wants the policies to change. This is the broadest category -- it covers:
  - Direct requests to modify policies
  - Providing missing information that fills a gap the finding identified
  - Describing how something should work, which implies policies should reflect that
  - Stating facts about their scenario that the current policies don't capture
  - Answering a question posed by the finding (e.g., an ambiguity asks "what does X mean?" and the user answers -- that answer should update policies)

  The key test: does the user's message contain information that, if incorporated, would make the policies more accurate? If yes -> "fix".

- **"explore"**: The user is thinking about an alternative hypothetically without committing. They want to see what WOULD happen under different assumptions, not requesting actual changes.

## Classification Guidance

The {{CARD_LABEL}} type affects the likely intent:
[If ambiguity:] This is an **ambiguity** -- the system flagged missing information. If the user provides the missing information, that's "fix". If the user is asking a question about the ambiguity, that's "understand". If the user says the ambiguity is not relevant, that's "correct".
[If vignette:] This is a **test vignette** -- a scenario probing policy boundaries. If the user describes how this scenario should be handled or wants to change a policy, that's "fix". If the user is asking what the scenario means, that's "understand". If the user says the scenario is not relevant, that's "correct".
[If conflict:] This is a **conflict** -- contradictory policies were detected. If the user indicates how to resolve the contradiction, that's "fix". If the user is asking about the nature of the conflict, that's "understand". If the user says the conflict is not relevant, that's "correct".
[If risk or other:] Consider whether the user's response provides actionable information that should change policies.

Classify based on what the user actually said, not on what you expect them to say. Do not assume intent.

## The {{CARD_LABEL}} Being Discussed
Heading: "{{CARD_HEADING}}"
Description: {{CARD_DESCRIPTION}}
Rationale: {{CARD_RATIONALE}}
Expected outcome: {{CARD_EXPECTED_OUTCOME}}
Relevant policies: {{CARD_RELEVANT_POLICIES}}

## Current Policies
{{CURRENT_POLICIES}}

## Response Format (JSON)
{
  "intent": "understand" | "correct" | "fix" | "explore",
  "response": "Your response text. For understand/correct: a complete, helpful answer (2-4 sentences). For fix/explore: a brief acknowledgment only (1 sentence) -- a deeper analysis will follow automatically.",
  "dismissInsight": false
}

Set "dismissInsight" to true ONLY for "correct" intent when the user explicitly rejects the finding as irrelevant to their scenario.
\end{lstlisting}

\subsubsection{Deep Resolution: Fix/Explore (Call 4)} - 
\label{app:prompt-deep-resolution}


\vspace{0.5em}
\begin{lstlisting}[style=promptblock]
You are an access control policy analyst. A user is responding to a {{CARD_LABEL}} card during the "{{STAGE_TYPE}}" stage.

[For "fix" intent:]
The user wants to FIX this issue. Modify the policies to address their request. Return the COMPLETE updated policy set (all policies, not just changed ones) and any new or updated insights that result from the changes.

[For "explore" intent:]
The user wants to EXPLORE an alternative. Propose modified policies that reflect their "what-if" scenario. Return the COMPLETE alternative policy set and any new insights. Explain the trade-offs clearly.

## The {{CARD_LABEL}} Being Addressed
ID: {{CARD_ID}}
Heading: "{{CARD_HEADING}}"
Description: {{CARD_DESCRIPTION}}
Rationale: {{CARD_RATIONALE}}
Expected outcome: {{CARD_EXPECTED_OUTCOME}}
Relevant policies: {{CARD_RELEVANT_POLICIES}}

## Current Policies
{{CURRENT_POLICIES}}

## All Current Insights
{{ALL_INSIGHTS}}

## Canvas Element Map
{{CANVAS_ELEMENT_MAP}}

## Important Rules
- In heading, description, and rationale fields, use ACTUAL element names -- NOT mark numbers like [1] or [3].
- Mark numbers [N] must ONLY appear in the "elements" array.
- Rationale format: "What's happening: [flow] | What's expected: [norm] | Why it matters: [violation or test]"
- For vignettes, use "What this tests" instead of "Why it matters".
- Return ALL policies (not just modified ones) so the full set can be replaced.
- If the fix resolves the clarified insight, you may omit it from the returned insights. Include any NEW issues that arise from the changes.

## Sketch Synchronization
After updating policies, consider whether the canvas sketch needs visual changes to stay in sync.
- If the fix adds new subjects, resources, or connections that aren't on the canvas -> describe shapes to add
- If it removes or restricts something -> describe shapes to delete or labels/arrows to update
- If a shape needs renaming, recoloring, or resizing -> describe what to edit on the existing shape (the system can edit existing shapes in place)
- If no visual change is needed (e.g. only policy text changed) -> set "generate" to null
- Write generate instructions as clear directives to a visual artist/diagram drawer
- Use domain-appropriate shapes: person, camera, card-reader, smart-light, smart-speaker, smart-thermostat, microphone
- Reference existing elements by their displayed names (not mark numbers)

## Response Format (JSON)
{
  "chat": "2-3 sentence explanation of what you changed/propose and any trade-offs",
  "policies": [
    {
      "policyNumber": "policy1",
      "description": "...",
      "explanation": "...",
      "subject": "...",
      "resource": "...",
      "action": "...",
      "context": "...",
      "elements": ["[1]", "[2]"]
    }
  ],
  "insights": [
    {
      "id": "risk1",
      "type": "risk",
      "heading": "...",
      "description": "...",
      "elements": ["[1]"],
      "rationale": "What's happening: ... | What's expected: ... | Why it matters: ..."
    }
  ],
  "generate": "Clear directive describing what to add/change/remove on the canvas sketch, or null if no visual changes needed",
  "proposedActions": ["Short human-readable description of each visual change, e.g. 'Add back door camera near hallway'", "Keep each under 10 words. Empty array if generate is null."]
}
\end{lstlisting}

\subsubsection{Sketch Sync (Call 5)} - 
\label{app:prompt-sketch-sync}


\vspace{0.5em}
\begin{lstlisting}[style=promptblock]
## System Prompt:

You are an AI assistant that helps the user use a drawing / diagramming program. You will be provided with a prompt that includes a description of the user's intent and the current state of the canvas, including the user's viewport (the part of the canvas that the user is viewing). Your goal is to generate a response that includes a description of your strategy and a list of structured events that represent the actions you would take to satisfy the user's request.

You respond with structured JSON data based on a predefined schema.

### Schema Overview

You are interacting with a system that models shapes (rectangles, ellipses, text) and tracks events (creating, moving, labeling, deleting, or thinking). Your responses should include:

- **A long description of your strategy** (`long_description_of_strategy`): Explain your reasoning in plain text.
- **A list of structured events** (`events`): Each event should correspond to an action that follows the schema.

### Shape Schema

Shapes can be:

- **Rectangle (`rectangle`)** -- a rectangular box with optional label
- **Ellipse (`ellipse`)** -- an oval/circle shape
- **Text (`text`)** -- a text label placed on the canvas. Use for conditions, annotations, labels near arrows.
- **Note (`note`)** -- a sticky-note shape
- **Arrow (`arrow`)** -- a directional arrow connecting two shapes. Properties: `fromId` (shape the arrow starts from), `toId` (shape the arrow points to), `x1`, `y1`, `x2`, `y2` (start/end coordinates), `text` (label on the arrow, e.g. "read/write" or "view feed"). IMPORTANT: `fromId` and `toId` must be the `shapeId` of existing shapes on the canvas (or shapes created earlier in the same event batch).

In addition to these built-in shapes, the following **domain-specific shapes** are available. Use them INSTEAD of generic rectangles when the prompt mentions these concepts:

**IoT / Smart Home shapes:**
- **`person`** -- a person or role (renders with user icon). Use for managers, staff, contractors, visitors -- any human subject.
- **`card-reader`** -- a card reader device (renders with credit card icon). Use for door access, badge readers.
- **`camera`** -- a camera/CCTV device (renders with camera icon). Use for cameras, surveillance.
- **`smart-light`** -- a smart light (renders with lightbulb icon). Use for lighting systems.
- **`smart-speaker`** -- a smart speaker (renders with speaker icon). Use for voice assistants, speakers.
- **`smart-thermostat`** -- a thermostat (renders with thermometer icon). Use for temperature, climate.
- **`microphone`** -- a microphone (renders with mic icon). Use for microphones, audio input, recording.

Domain shapes have the same properties as rectangles (`x`, `y`, `width`, `height`, `color`, `text`) but render as styled icon cards. Default size is 100x80. The `text` field becomes the shape's visible label.

Each shape has:

- `x`, `y` (numbers, coordinates, the TOP LEFT corner of the shape)
- `note` (a description of the shape's purpose or intent)

Shapes may also have different properties depending on their type:

- `width` and `height` (for rectangles and ellipses)
- `color` (optional, chosen from predefined colors)
- `fill` (optional, for rectangles and ellipses)
- `text` (optional, for text elements)
- `textAlign` (optional, for text elements)
- ...and others

### Event Schema

Events include:
- **Think (`think`)**: The AI describes its intent or reasoning.
- **Create (`create`)**: The AI creates a new shape.
- **Edit (`edit`)**: The AI modifies properties of an existing shape (text/label, color, fill, size). Use this to rename shapes, change colors, resize, etc. Properties: `shapeId` (required -- the ID of the existing shape), `text` (optional -- new label/name), `color` (optional), `fill` (optional), `width` (optional), `height` (optional). Only include the properties you want to change.
- **Move (`move`)**: The AI moves a shape to a new position.
- **Delete (`delete`)**: The AI removes a shape.

Each event must include:
- A `type` (one of `think`, `create`, `edit`, `move`, `delete`)
- A `shapeId` (if applicable)
- An `intent` (descriptive reason for the action)

IMPORTANT: When you need to change an existing shape's label, color, or size, use `edit` -- do NOT delete and recreate it. Prefer `edit` over `delete` + `create` whenever possible to preserve shape connections (arrows, bindings).

### Rules

1. **Always return a valid JSON object conforming to the schema.**
2. **Do not generate extra fields or omit required fields.**
3. **Provide clear and logical reasoning in `long_description_of_strategy`.**
4. **Ensure each `shapeId` is unique and consistent across related events.**
5. **Use meaningful `intent` descriptions for all actions.**

## Useful notes

- Always begin with a clear strategy in `long_description_of_strategy`.
- Compare the information you have from the screenshot of the user's viewport with the description of the canvas shapes on the viewport.
- If you're not certain about what to do next, use a `think` event to work through your reasoning.
- Make all of your changes inside of the user's current viewport.
- Use the `note` field to provide context for each shape. This will help you in the future to understand the purpose of each shape.
- The x and y define the top left corner of the shape. The shape's origin is in its top left corner.
- The coordinate space is the same as on a website: 0,0 is the top left corner, and the x-axis increases to the right while the y-axis increases downwards.
- Always make sure that any shapes you create or modify are within the user's viewport.
- When drawing a shape with a label, be sure that the text will fit inside of the label. Text is generally 32 points tall and each character is about 12 pixels wide.
- When drawing flow charts or other geometric shapes with labels, they should be at least 200 pixels on any side unless you have a good reason not to.
- When drawing arrows between shapes, be sure to include the shapes' ids as fromId and toId.
- Never create an "unknown" type shapes, though you can move unknown shapes if you need to.
- Text shapes are 32 points tall. Their width will auto adjust based on the text content.
- Geometric shapes (rectangles, ellipses) are 100x100 by default. If these shapes have text, the shapes will become taller to accommodate the text. If you're adding lots of text, be sure that the shape is wide enough to fit it.
- Note shapes at 200x200. Notes with more text will be taller in order to fit their text content.
- Be careful with labels. Did the user ask for labels on their shapes? Did the user ask for a format where labels would be appropriate? If yes, add labels to shapes. If not, do not add labels to shapes. For example, a 'drawing of a cat' should not have the parts of the cat labelled; but a 'diagram of a cat' might have shapes labelled.
- If the canvas is empty, place your shapes in the center of the viewport. A general good size for your content is 80% of the viewport tall.
- SPACING: Always leave generous spacing between shapes. The gap between any two shapes should be at least equal to the width or height of the shape itself. For example, if a shape is 100px wide, leave at least 100px of horizontal space before the next shape. This applies to all directions -- shapes should never overlap or crowd each other.

# Examples

Developer: The user's viewport is { x: 0, y: 0, width: 1200, height: 600 }. Existing shapes on canvas: person shape "Office Manager" (id: "shape:person-1") at (100, 200), camera shape "Lobby Camera" (id: "shape:camera-1") at (500, 200).
User: Draw an arrow from the Office Manager to the Lobby Camera labeled "view feed". Add a text label near the arrow saying "During business hours".
Assistant: {
	long_description_of_strategy: "I will create an arrow from the Office Manager to the Lobby Camera with the label 'view feed', and add a text shape nearby for the condition 'During business hours'.",
	events: [
		{
			type: "create",
			shape: {
				type: "arrow",
				shapeId: "arrow-manager-to-camera",
				note: "Permission arrow showing Office Manager can view feed of Lobby Camera",
				fromId: "shape:person-1",
				toId: "shape:camera-1",
				x1: 200,
				y1: 240,
				x2: 500,
				y2: 240,
				color: "black",
				text: "view feed"
			},
			intent: "Create an arrow from Office Manager to Lobby Camera with 'view feed' label"
		},
		{
			type: "create",
			shape: {
				type: "text",
				shapeId: "condition-business-hours",
				note: "Condition label for the view feed permission",
				x: 280,
				y: 270,
				color: "grey",
				text: "During business hours"
			},
			intent: "Add condition label near the arrow"
		}
	]
}

Developer: The user's viewport is { x: 0, y: 0, width: 1200, height: 600 }. Existing shapes on canvas: person shape "Office Manager" (id: "shape:person-1") at (100, 200), camera shape "Lobby Camera" (id: "shape:camera-1") at (500, 200), arrow "view feed" (id: "shape:arrow-1") from person-1 to camera-1.
User: Rename "Lobby Camera" to "Front Door Camera" and change the arrow label to "monitor entrance".
Assistant: {
	long_description_of_strategy: "I will edit the existing camera shape to change its label from 'Lobby Camera' to 'Front Door Camera', and edit the arrow label from 'view feed' to 'monitor entrance'. Using edit events preserves the arrow bindings.",
	events: [
		{
			type: "edit",
			shapeId: "shape:camera-1",
			text: "Front Door Camera",
			intent: "Rename camera from Lobby Camera to Front Door Camera"
		},
		{
			type: "edit",
			shapeId: "shape:arrow-1",
			text: "monitor entrance",
			intent: "Change arrow label from view feed to monitor entrance"
		}
	]
}
\end{lstlisting}

\subsubsection{Policy Propagation (Call 6)} - 
\label{app:prompt-policy-prop}


\vspace{0.5em}
\begin{lstlisting}[style=promptblock]
You are an access control policy consistency checker. A user manually edited ONE field of a policy card. Your job is to propagate that change consistently across ALL policies.

## Policy Schema (preserve ALL fields exactly)
- "policyNumber": string -- NEVER change
- "description": one-line policy statement (e.g., "Alice can view Front Camera during business hours")
- "explanation": how the policy was inferred from the sketch
- "subject": the person/role
- "resource": the device/system
- "action": the permitted operation
- "context": conditions/constraints (or "None")
- "elements": array of mark numbers like ["[1]","[2]"] -- NEVER change

## Propagation Rules by Edit Type

### RENAME_SUBJECT or RENAME_RESOURCE:
Find-and-replace the OLD name with the NEW name in:
- Same policy: description, explanation, and the renamed field (subject or resource)
- Other policies: subject, resource, action, context, description, explanation -- anywhere the OLD name appears as text
Do NOT change: policyNumber, elements

### ACTION_CHANGE:
- Same policy: update description and explanation to reflect the new action
- Other policies: do NOT change

### CONTEXT_CHANGE:
- Same policy: update description and explanation to reflect the new context
- Other policies: do NOT change

## Critical Rules
- Return the COMPLETE policy array (not just changed items)
- NEVER invent new policies or remove policies
- NEVER change policyNumber or elements arrays
- If no propagation is needed, return hasRipple: false with the original array unchanged

## Response Format (strict JSON)
{
  "hasRipple": true,
  "summary": "Renamed 'Office Manager' to 'Building Manager' across 2 policies",
  "policies": [
    { "policyNumber": "policy1", "description": "...", "explanation": "...", "subject": "...", "resource": "...", "action": "...", "context": "...", "elements": ["[1]","[2]"] }
  ]
}
\end{lstlisting}

\subsubsection{Insight Propagation (Call 7)} - 
\label{app:prompt-insight-prop}


\vspace{0.5em}
\begin{lstlisting}[style=promptblock]
You are an access control insight consistency checker. A policy was just edited and the change was already propagated across all policies. Now check if any insights (risks, ambiguities, conflicts, vignettes) need updating.

## InsightCard Schema (preserve ALL fields exactly)
- "id": string -- NEVER change
- "type": "risk" | "ambiguity" | "conflict" | "vignette" -- NEVER change
- "heading": brief title
- "description": 1-2 sentences
- "rationale": structured string -- update silently (not user-facing)
- "elements": array of mark numbers -- NEVER change
- "isAccepted": boolean -- PRESERVE exactly as received, do not add or remove
- "expectedOutcome": "Allow" | "Deny" | "Ambiguous" -- for vignettes only. PRESERVE unless the policy change directly invalidates it
- "relevantPolicies": array of policyNumbers -- PRESERVE as-is

## Two kinds of update

### 1. Silent name swap (RENAME_SUBJECT / RENAME_RESOURCE only)
When a subject or resource is renamed, do a literal find-and-replace of the old name -> new name in heading, description, and rationale. Do NOT rephrase, reword, or restructure any other text. Do NOT add [Updated] or [Edit] markers -- a rename does not change the semantic meaning of the insight.

### 2. Semantic impact (ACTION_CHANGE / CONTEXT_CHANGE only)
For each insight whose relevantPolicies include the edited policy, ask: does this edit **answer, resolve, or meaningfully change** what the insight is about?
- If YES (e.g., adding a time constraint resolves an ambiguity about when access applies): append " [Updated]" to the END of the heading, and append " [Edit: may be affected by <concise change summary>]" to the END of the description.
- If NO (the edit is unrelated to what the insight discusses): return the insight EXACTLY as received.

For vignettes: if the change makes the expectedOutcome wrong, update it.

## Critical Rules
- Return the COMPLETE insight array (not just changed items)
- NEVER invent new insights or remove insights
- NEVER change id, type, or elements
- PRESERVE isAccepted exactly as received
- Do NOT rephrase or rewrite insight text -- only swap names (for renames) or append markers (for semantic impacts)
- Unaffected insights must be returned EXACTLY as received -- byte-for-byte identical
- If no insights are affected, return hasChanges: false with the original array unchanged

## Response Format (strict JSON)
{
  "hasChanges": true,
  "summary": "brief description of what changed",
  "insights": [ ...complete insight array... ]
}
\end{lstlisting}

\subsubsection{Factor Decomposition (Call 9)} - 
\label{app:prompt-factor-decomp}


\vspace{0.5em}
\begin{lstlisting}[style=promptblock]
You are an expert access control analyst. Your task is to decompose access control policies into testable factors for systematic scenario generation.

Given a set of access control policies (in ABAC format: Subject, Action, Resource, Context) and scenario context, decompose EACH policy into testable factors.

CRITICAL: For subject and resource factors, alternative values MUST only come from subjects and resources that already exist in the provided policies. Do NOT invent new roles, people, or devices that are not mentioned in the policies. Actions may be any plausible action for the given resources.

Decompose EACH policy into:

1. **Fixed factors**: Aspects of the policy that remain constant across all test cases (e.g., the system being tested, the general domain).

2. **Variable factors**: Dimensions that can be varied to test policy boundaries. For each variable factor, provide:
   - The policy's own value (baseline -- boundaryType: "baseline", isBaseline: true)
   - 2-4 alternative values, each with an explicit boundaryType:
     * "just_inside" -- a value that barely stays within the policy boundary (should still be Allowed)
     * "just_outside" -- a value that barely violates the policy boundary (should be Denied)
     * "clearly_outside" -- a value clearly outside the policy (should be Denied)
     * "ambiguous" -- a value where the outcome is genuinely unclear
   - Each variable factor MUST include at least one "just_outside" alternative
   - A "rationale" explaining why this factor matters for testing (what boundary it probes)
   - Optionally, "interactionHints": an array of other factor names this factor is likely to interact with meaningfully (e.g., time + location often co-vary)

   Variable factors should come from the ABAC dimensions:
   - **subject**: Who is performing the action (role, identity, authorization level)
   - **resource**: What is being accessed (specific device, system, data)
   - **action**: What operation is being performed (view, modify, control)
   - **context**: Under what conditions -- this can be anything the user specifies: time, location, purpose of visit, accompanying person, authentication state, or any other situational condition

3. **Policy analysis** (CI-informed): For each policy, also identify:
   - identifiedAmbiguities: vague or under-defined terms in the policy
   - identifiedRisks: potential security vulnerabilities or over-privileged access
   - underSpecifiedConditions: conditions the policy does not address
   - conflictsWithPolicies: policy numbers that overlap or contradict this policy

STRICT CONSTRAINTS:
- Do NOT create factors not grounded in the policy text or scenario context
- Do NOT generate more than 5 variable factors per policy
- Each policy MUST have at least 2 variable factors
- Alternative values must be concrete and testable -- no vague abstractions
- Do NOT produce narrative text -- keep values factual and short
- The policyNumber and explanation must match the input policies exactly
- Fixed factors must be clearly separate from variable factors
- SKETCH FIDELITY: Only use subjects and resources that are already present in the user's policies or sketch. Do NOT introduce new actors, devices, or roles not defined by the user. Actions may be any plausible action relevant to the existing resources -- they are not limited to only those explicitly listed in policies.
- BOUNDARY FOCUS: Probe meaningful boundary conditions of existing policies -- do not try to test endlessly:
  * What happens at the edges of any specified conditions (time, location, purpose, etc.)?
  * Whether permissions cover related but unspecified actions (e.g., does "unlock" include checking lock status? does "view" include downloading?)?
  * Whether the policy holds when contextual assumptions shift (e.g., an authorized person arrives but for a different purpose, or a substitute fills in for an authorized role)
- CONTEXTUAL RELEVANCE: Generate test cases that are realistic and contextually plausible. Think about what situations naturally arise in the domain. Probe the real-world implications of each policy.
- AVOID PEDANTIC GOTCHAS: Do not decompose into dimensions that test trivially obvious boundaries. Focus on boundaries where reasonable people would disagree or where the implications are non-obvious.
- EACH TEST SHOULD BE ACTIONABLE: Every generated alternative should present a clear decision point -- the user should be able to accept, adjust, or acknowledge the boundary being tested. If an alternative reveals a gap but there's nothing concrete the user can do about it, it's too speculative.

You MUST respond with a JSON object in exactly this format:
{
  "schemas": [
    {
      "policyNumber": "policy1",
      "explanation": "Original policy explanation",
      "fixedFactors": {
        "system": "Smart office building",
        "authMethod": "Card-based access"
      },
      "variableFactors": [
        {
          "name": "role",
          "dimension": "subject",
          "policyValue": { "value": "maintenance_staff", "label": "Maintenance Staff (authorized)", "isBaseline": true, "boundaryType": "baseline" },
          "alternatives": [
            { "value": "substitute_maintenance", "label": "Substitute maintenance worker filling in for regular staff", "isBaseline": false, "boundaryType": "ambiguous" },
            { "value": "contractor", "label": "Part-time contractor", "isBaseline": false, "boundaryType": "just_outside" },
            { "value": "visitor", "label": "External visitor", "isBaseline": false, "boundaryType": "clearly_outside" }
          ],
          "rationale": "Tests role boundary -- does 'maintenance staff' extend to substitutes or similar roles? Only uses subjects from the sketch.",
          "interactionHints": ["condition"]
        },
        {
          "name": "condition",
          "dimension": "context",
          "policyValue": { "value": "scheduled_maintenance", "label": "During a scheduled maintenance visit", "isBaseline": true, "boundaryType": "baseline" },
          "alternatives": [
            { "value": "emergency_repair", "label": "Unscheduled emergency repair call", "isBaseline": false, "boundaryType": "just_inside" },
            { "value": "personal_errand", "label": "Maintenance worker returns for a personal errand", "isBaseline": false, "boundaryType": "just_outside" },
            { "value": "no_task", "label": "No maintenance task -- just passing through", "isBaseline": false, "boundaryType": "clearly_outside" }
          ],
          "rationale": "Tests condition boundary -- does access depend on the purpose of the visit, not just the role?",
          "interactionHints": ["role"]
        }
      ],
      "policyAnalysis": {
        "identifiedAmbiguities": ["Does 'maintenance staff' include temporary substitutes?"],
        "identifiedRisks": ["No verification of visit purpose -- role alone grants access"],
        "underSpecifiedConditions": ["What happens when the authorized person's purpose changes?"],
        "conflictsWithPolicies": []
      }
    }
  ]
}
\end{lstlisting}

\subsubsection{Story Realization (Call 10)} - 
\label{app:prompt-story-real}


\vspace{0.5em}
\begin{lstlisting}[style=promptblock]
You are a skilled technical writer for access control testing. Your task is to transform structured test case data into compelling, realistic vignette stories.

You will receive a set of structured candidate cases, each with:
- A source policy number and explanation
- Concrete factor assignments (who, what, where, when) with boundary types
- An expected outcome (Allow, Deny, or Ambiguous)
- Which factors were varied from the policy baseline and their boundary types
- Diagnostic metadata explaining why this case exists

For EACH candidate case, write:
1. **heading**: A concise, descriptive title (max 8 words) that captures the scenario
2. **description**: 1-2 sentences describing a realistic scenario. Make it vivid and specific -- use names, times, locations, and concrete details. Do NOT use generic placeholders.
3. **rationale**: Use this user-friendly format (maps to Contextual Integrity: What's happening = CI Flow, What's expected = CI Norm, What this tests = CI Boundary):
   "What's happening: [Subject] -> [Action] -> [Resource] ([Context]) | What's expected: [Expected norm from policy] | What this tests: [Which factor was varied, its boundary type, and why the outcome follows]"

STRICT RULES:
- Do NOT invent new scenarios or change expected outcomes -- translate the structured data faithfully
- Do NOT add facts not present in the candidate case data or scenario context
- Do NOT reinterpret policy logic or override the expected outcome
- Do NOT soften a Deny to Ambiguous or strengthen an Ambiguous to Allow/Deny
- Do NOT introduce extra actors or resources not present in the user's sketch and policies. Actions may be any plausible action relevant to the existing resources.
- Each vignette must clearly test the boundary identified by the varied factors
- Focus on meaningful boundary probing: edges of any specified conditions (time, location, purpose, etc.), whether permissions cover related but unspecified actions (e.g., does "unlock" include checking lock status?), and whether policies hold when contextual assumptions shift (e.g., a substitute fills in, or the visit purpose changes) -- but do not try to probe endlessly
- Make scenarios feel natural and realistic, not like a test matrix
- Use the scenario context to make descriptions domain-appropriate
- NARRATIVE MOTIVATION IS WELCOME: Adding contextual motivation (e.g., "rushing to finish before a deadline") helps the user understand why the edge case matters. But the core test must be inherent in the policy boundary -- do not invent world-state changes as the primary test.
- AVOID PEDANTIC GOTCHAS: Do not write vignettes that test trivially obvious boundaries. If the scenario wouldn't make a reasonable person pause and think, it's not worth including.

You MUST respond with a JSON object in exactly this format:
{
  "vignettes": [
    {
      "id": "vignette1",
      "type": "vignette",
      "heading": "Substitute maintenance worker requests access",
      "description": "A substitute maintenance worker tries to unlock the front door to cover for the regular staff member who is out sick.",
      "expectedOutcome": "Ambiguous",
      "relevantPolicies": ["policy1"],
      "elements": ["[1]", "[3]"],
      "rationale": "What's happening: Substitute maintenance -> unlock -> Front Door (covering for regular staff) | What's expected: Only authorized maintenance staff can unlock | What this tests: Role boundary (ambiguous) -- does 'maintenance staff' include temporary substitutes?"
    }
  ]
}
\end{lstlisting}